\documentclass[twocolumn,letterpaper,aps,prd,superscriptaddress,showpacs,floatfix]{revtex4}

\usepackage{graphicx}
\usepackage{amsmath}
\usepackage{mathtools}
\usepackage{float}
\usepackage{xspace}

\begin{document}

\title{Cross sections and double-helicity asymmetries of midrapidity 
inclusive charged hadrons in $p$+$p$ collisions at $\sqrt{s}= 62.4$ GeV}

\newcommand{\abilene}{Abilene Christian University, Abilene, Texas 79699, USA}
\newcommand{\acadsin}{Institute of Physics, Academia Sinica, Taipei 11529, Taiwan}
\newcommand{\banaras}{Department of Physics, Banaras Hindu University, Varanasi 221005, India}
\newcommand{\barc}{Bhabha Atomic Research Centre, Bombay 400 085, India}
\newcommand{\bnlcoll}{Collider-Accelerator Department, Brookhaven National Laboratory, Upton, New York 11973-5000, USA}
\newcommand{\bnlphys}{Physics Department, Brookhaven National Laboratory, Upton, New York 11973-5000, USA}
\newcommand{\caucr}{University of California - Riverside, Riverside, California 92521, USA}
\newcommand{\charlesczech}{Charles University, Ovocn\'{y} trh 5, Praha 1, 116 36, Prague, Czech Republic}
\newcommand{\ciae}{Science and Technology on Nuclear Data Laboratory, China Institute of Atomic Energy, Beijing 102413, P.~R.~China}
\newcommand{\cns}{Center for Nuclear Study, Graduate School of Science, University of Tokyo, 7-3-1 Hongo, Bunkyo, Tokyo 113-0033, Japan}
\newcommand{\colorado}{University of Colorado, Boulder, Colorado 80309, USA}
\newcommand{\columbia}{Columbia University, New York, New York 10027 and Nevis Laboratories, Irvington, New York 10533, USA}
\newcommand{\czechtech}{Czech Technical University, Zikova 4, 166 36 Prague 6, Czech Republic}
\newcommand{\dapnia}{Dapnia, CEA Saclay, F-91191, Gif-sur-Yvette, France}
\newcommand{\debrecen}{Debrecen University, H-4010 Debrecen, Egyetem t{\'e}r 1, Hungary}
\newcommand{\elte}{ELTE, E{\"o}tv{\"o}s Lor{\'a}nd University, H - 1117 Budapest, P{\'a}zm{\'a}ny P. s. 1/A, Hungary}
\newcommand{\fit}{Florida Institute of Technology, Melbourne, Florida 32901, USA}
\newcommand{\fsu}{Florida State University, Tallahassee, Florida 32306, USA}
\newcommand{\gsu}{Georgia State University, Atlanta, Georgia 30303, USA}
\newcommand{\hiroshima}{Hiroshima University, Kagamiyama, Higashi-Hiroshima 739-8526, Japan}
\newcommand{\ihepprot}{IHEP Protvino, State Research Center of Russian Federation, Institute for High Energy Physics, Protvino, 142281, Russia}
\newcommand{\illuiuc}{University of Illinois at Urbana-Champaign, Urbana, Illinois 61801, USA}
\newcommand{\inrras}{Institute for Nuclear Research of the Russian Academy of Sciences, prospekt 60-letiya Oktyabrya 7a, Moscow 117312, Russia}
\newcommand{\instpasczech}{Institute of Physics, Academy of Sciences of the Czech Republic, Na Slovance 2, 182 21 Prague 8, Czech Republic}
\newcommand{\isu}{Iowa State University, Ames, Iowa 50011, USA}
\newcommand{\jinrdubna}{Joint Institute for Nuclear Research, 141980 Dubna, Moscow Region, Russia}
\newcommand{\kek}{KEK, High Energy Accelerator Research Organization, Tsukuba, Ibaraki 305-0801, Japan}
\newcommand{\korea}{Korea University, Seoul, 136-701, Korea}
\newcommand{\kurchatov}{Russian Research Center ``Kurchatov Institute", Moscow, 123098 Russia}
\newcommand{\kyoto}{Kyoto University, Kyoto 606-8502, Japan}
\newcommand{\labllr}{Laboratoire Leprince-Ringuet, Ecole Polytechnique, CNRS-IN2P3, Route de Saclay, F-91128, Palaiseau, France}
\newcommand{\lawllnl}{Lawrence Livermore National Laboratory, Livermore, California 94550, USA}
\newcommand{\losalamos}{Los Alamos National Laboratory, Los Alamos, New Mexico 87545, USA}
\newcommand{\lpc}{LPC, Universit{\'e} Blaise Pascal, CNRS-IN2P3, Clermont-Fd, 63177 Aubiere Cedex, France}
\newcommand{\lund}{Department of Physics, Lund University, Box 118, SE-221 00 Lund, Sweden}
\newcommand{\mass}{Department of Physics, University of Massachusetts, Amherst, Massachusetts 01003-9337, USA }
\newcommand{\muenster}{Institut f\"ur Kernphysik, University of Muenster, D-48149 Muenster, Germany}
\newcommand{\muhlenberg}{Muhlenberg College, Allentown, Pennsylvania 18104-5586, USA}
\newcommand{\myongji}{Myongji University, Yongin, Kyonggido 449-728, Korea}
\newcommand{\nagasaki}{Nagasaki Institute of Applied Science, Nagasaki-shi, Nagasaki 851-0193, Japan}
\newcommand{\newmex}{University of New Mexico, Albuquerque, New Mexico 87131, USA }
\newcommand{\nmsu}{New Mexico State University, Las Cruces, New Mexico 88003, USA}
\newcommand{\ornl}{Oak Ridge National Laboratory, Oak Ridge, Tennessee 37831, USA}
\newcommand{\orsay}{IPN-Orsay, Universite Paris Sud, CNRS-IN2P3, BP1, F-91406, Orsay, France}
\newcommand{\peking}{Peking University, Beijing 100871, P.~R.~China}
\newcommand{\pnpi}{PNPI, Petersburg Nuclear Physics Institute, Gatchina, Leningrad region, 188300, Russia}
\newcommand{\riken}{RIKEN Nishina Center for Accelerator-Based Science, Wako, Saitama 351-0198, Japan}
\newcommand{\rikjrbrc}{RIKEN BNL Research Center, Brookhaven National Laboratory, Upton, New York 11973-5000, USA}
\newcommand{\rikkyo}{Physics Department, Rikkyo University, 3-34-1 Nishi-Ikebukuro, Toshima, Tokyo 171-8501, Japan}
\newcommand{\saispbstu}{Saint Petersburg State Polytechnic University, St. Petersburg, 195251 Russia}
\newcommand{\saopaulo}{Universidade de S{\~a}o Paulo, Instituto de F\'{\i}sica, Caixa Postal 66318, S{\~a}o Paulo CEP05315-970, Brazil}
\newcommand{\seoulnat}{Seoul National University, Seoul, Korea}
\newcommand{\stonybrkc}{Chemistry Department, Stony Brook University, SUNY, Stony Brook, New York 11794-3400, USA}
\newcommand{\stonycrkp}{Department of Physics and Astronomy, Stony Brook University, SUNY, Stony Brook, New York 11794-3400, USA}
\newcommand{\subatech}{SUBATECH (Ecole des Mines de Nantes, CNRS-IN2P3, Universit{\'e} de Nantes) BP 20722 - 44307, Nantes, France}
\newcommand{\tenn}{University of Tennessee, Knoxville, Tennessee 37996, USA}
\newcommand{\titech}{Department of Physics, Tokyo Institute of Technology, Oh-okayama, Meguro, Tokyo 152-8551, Japan}
\newcommand{\tsukuba}{Institute of Physics, University of Tsukuba, Tsukuba, Ibaraki 305, Japan}
\newcommand{\vandy}{Vanderbilt University, Nashville, Tennessee 37235, USA}
\newcommand{\waseda}{Waseda University, Advanced Research Institute for Science and Engineering, 17 Kikui-cho, Shinjuku-ku, Tokyo 162-0044, Japan}
\newcommand{\weizmann}{Weizmann Institute, Rehovot 76100, Israel}
\newcommand{\wigner}{Institute for Particle and Nuclear Physics, Wigner Research Centre for Physics, Hungarian Academy of Sciences (Wigner RCP, RMKI), H-1525 Budapest 114, POBox 49, Budapest, Hungary}
\newcommand{\yonsei}{Yonsei University, IPAP, Seoul 120-749, Korea}
\affiliation{\abilene}
\affiliation{\acadsin}
\affiliation{\banaras}
\affiliation{\barc}
\affiliation{\bnlcoll}
\affiliation{\bnlphys}
\affiliation{\caucr}
\affiliation{\charlesczech}
\affiliation{\ciae}
\affiliation{\cns}
\affiliation{\colorado}
\affiliation{\columbia}
\affiliation{\czechtech}
\affiliation{\dapnia}
\affiliation{\debrecen}
\affiliation{\elte}
\affiliation{\fit}
\affiliation{\fsu}
\affiliation{\gsu}
\affiliation{\hiroshima}
\affiliation{\ihepprot}
\affiliation{\illuiuc}
\affiliation{\inrras}
\affiliation{\instpasczech}
\affiliation{\isu}
\affiliation{\jinrdubna}
\affiliation{\kek}
\affiliation{\korea}
\affiliation{\kurchatov}
\affiliation{\kyoto}
\affiliation{\labllr}
\affiliation{\lawllnl}
\affiliation{\losalamos}
\affiliation{\lpc}
\affiliation{\lund}
\affiliation{\mass}
\affiliation{\muenster}
\affiliation{\muhlenberg}
\affiliation{\myongji}
\affiliation{\nagasaki}
\affiliation{\newmex}
\affiliation{\nmsu}
\affiliation{\ornl}
\affiliation{\orsay}
\affiliation{\peking}
\affiliation{\pnpi}
\affiliation{\riken}
\affiliation{\rikjrbrc}
\affiliation{\rikkyo}
\affiliation{\saispbstu}
\affiliation{\saopaulo}
\affiliation{\seoulnat}
\affiliation{\stonybrkc}
\affiliation{\stonycrkp}
\affiliation{\subatech}
\affiliation{\tenn}
\affiliation{\titech}
\affiliation{\tsukuba}
\affiliation{\vandy}
\affiliation{\waseda}
\affiliation{\weizmann}
\affiliation{\wigner}
\affiliation{\yonsei}
\author{A.~Adare} \affiliation{\colorado}
\author{S.~Afanasiev} \affiliation{\jinrdubna}
\author{C.~Aidala} \affiliation{\mass}
\author{N.N.~Ajitanand} \affiliation{\stonybrkc}
\author{Y.~Akiba} \affiliation{\riken} \affiliation{\rikjrbrc}
\author{H.~Al-Bataineh} \affiliation{\nmsu}
\author{J.~Alexander} \affiliation{\stonybrkc}
\author{K.~Aoki} \affiliation{\kyoto} \affiliation{\riken}
\author{L.~Aphecetche} \affiliation{\subatech}
\author{J.~Asai} \affiliation{\riken}
% ...
\collaboration{PHENIX Collaboration} \noaffiliation
\date{\today}

%-----------------------
\begin{abstract}

Unpolarized cross sections and double-helicity asymmetries of 
single-inclusive positive and negative charged hadrons at 
midrapidity from $p$+$p$ collisions at $\sqrt{s} = 62.4$~GeV 
are presented.  The PHENIX measurement of the cross sections 
for $1.0 < p_T < 4.5$ GeV/$c$ are consistent with perturbative 
QCD calculations at next-to-leading order in the 
strong-coupling constant, $\alpha_s$.  Resummed pQCD 
calculations including terms with next-to-leading-log 
accuracy, yielding reduced theoretical uncertainties, also 
agree with the data. The double-helicity asymmetry, sensitive 
at leading order to the gluon polarization in a momentum-fraction 
range of $0.05 \lesssim x_{gluon} \lesssim 0.2$, is consistent 
with recent global parameterizations disfavoring 
large gluon polarization.

\end{abstract}


\maketitle

%-----------------------

\section{Introduction}

The comparison of cross-section predictions with data on 
single-inclusive hadron production in hadronic collisions, 
$p$+$p \rightarrow h+X$, is important for understanding perturbative 
quantum chromodynamics (pQCD).  For hadrons produced with transverse 
momenta $p_{T} \gg \Lambda_{QCD}$, the cross section factorizes into a 
convolution involving long-distance and short-distance 
components~\cite{FactPropose, FactProof}.  Long-distance components 
include universal parton-distribution functions (PDFs) describing the 
partonic structure of the initial hadrons and fragmentation functions 
(FFs) for the final-state hadron.  The short-distance part describes the 
hard scattering of partons.  The long-distance components, PDFs and 
FFs, can be extracted from other processes, such as deep-inelastic 
scattering and hadron production in $e^{+}e^{-}$ colliders.  This allows 
for a test of the short-distance part of the convolution, which can be 
estimated using pQCD.  In particular, differences between data and 
predictions can indicate the importance of neglected higher-order terms 
in the expansion or power-suppressed contributions~\cite{Resum1}.

Next-to-leading-order (NLO) pQCD and collinear factorization 
successfully describe RHIC cross-section measurements at a 
center-of-mass energy ($\sqrt{s}$) of 200~GeV for midrapidity neutral 
pions~\cite{PI0200, STAR3}, jets~\cite{PHENIXJet, PHENIX3, STAR1}, and 
direct photons~\cite{PHENIX4}, as well as forward rapidity pions and 
kaons~\cite{STAR2, BRAHMS}.  However, at lower $\sqrt{s}$, in 
particular in fixed-target experiments with $20 \lesssim \sqrt{s} 
\lesssim 40$ GeV, NLO pQCD calculations significantly underpredict 
hadron production, by factors of three or more~\cite{Resum1}. The 
consistency between NLO estimations and data at low $\sqrt{s}$ was 
improved \cite{Resum1,Resum,Resum2} by including the resummation of 
large logarithmic corrections to the partonic cross section to all 
orders in the strong coupling $\alpha_{s}$. The corrections are of the 
form $\alpha_{s}^{k}\ln^{2k}\left(1-\hat{x}^{2}_{T}\right)$ for the 
$k$th-order term in the perturbative expansion. Here $\hat{x}_{T}\equiv 
2\hat{p}_{T}/\sqrt{\hat{s}}$, where $\hat{p}_{T}=p_{T}/z$ is the 
transverse momentum of the parton fragmenting into the observed hadron 
with a fraction $z$ of the parton transverse momentum, and 
$\sqrt{\hat{s}} = \sqrt{x_1x_2s}$ is the partonic center-of-mass energy 
where $x_1,x_2$ are momentum fractions carried by two interacting 
partons. The corrections are especially relevant in the threshold 
regime $\hat{x}_{T}\rightarrow 1$ in which the initial partons have 
just enough energy to produce a high-transverse-momentum parton 
fragmenting into the observed hadron. In this regime gluon 
Bremsstrahlung is suppressed, and these corrections are 
large~\cite{Resum}. However, the addition of the resummed 
next-to-leading-log (NLL) terms to an NLO calculation may not provide 
the best means of describing data in a given kinematic region, for 
example, when the (unknown) higher-order terms that are omitted from 
the calculation have comparable magnitude and opposite sign to the NLL 
terms.  It is therefore important to test pQCD calculations against 
data in a region of intermediate $\sqrt{s}$, to better define the 
kinematic ranges over which pQCD calculations can be applied with 
confidence.

The PHENIX-dectector data presented here for nonidentified 
charged-hadron production make use of approximately five times the acceptance and 
different detection techniques compared to our previous 
identified-charged-hadron analysis~\cite{PH62idch}.
The new results allow tests of NLO and NLL 
predictions based on an independent measurement.  In addition, the 
theoretical calculations make use of different fragmentation functions 
than for identified particle calculations.  Alternatively, assuming the 
reliability of the short-distance aspects of the theory, the data may 
be used to refine our knowledge of fragmentation functions.  Present 
measurements also cover a greater $p_T$ range than the identified 
charged-hadron cross sections, where the measured momentum ranges for 
pions, kaons, and protons are 0.3--3 GeV/$c$, 0.4--2 GeV/$c$ and 
0.5--4.5 GeV/$c$ respectively~\cite{PH62idch}. These cross-section 
measurements of nonidentified charged hadrons (combinations of 
$\pi^{\pm}$, $K^{\pm}$, $p^{\pm}$) are also important as baselines for 
extracting nuclear modification factors in high-$p_{T}$ hadron 
production in heavy ion collisions at RHIC~\cite{PHENIX_heavydAu, 
PHENIX_heavyAuAu}.

The charged hadrons in these measurements were produced from collisions 
of transversely- and longitudinally-polarized proton beams, a unique 
capability of RHIC~\cite{RHIC_polpp}.  While the cross-section 
measurements discussed above require averaging over the beam 
polarizations, sorting the hadron yields by colliding proton helicities 
(for longitudinal beam polarizations) provides sensitivity to the 
helicity PDFs~\cite{BunceAnnRev}. The ability to probe helicity PDFs is 
essential for understanding the spin structure of the 
proton~\cite{SPINRev}.

The spin of the proton originates from the spin and orbital angular 
momenta of its quark, antiquark, and gluon constituents. The 
contribution carried by quark and antiquark spin, determined from 
polarized deep-inelastic scattering (pDIS) 
experiments~\cite{EMC,E142n,E154n,HERMESn,E143pd,SMCpd,E155p,JLABn,JLABp,COMPASSd,HERMESpd} 
using polarized leptons and polarized nucleons, is $\sim 
25\%$~\cite{SPINRev,GRSV,BB,LSS2010}. This is surprisingly 
small~\cite{SPINRev} and implies that the majority of the spin of the 
proton must originate from gluon spin and/or orbital angular momentum.

Colliding longitudinally-polarized proton beams provides sensitivity to 
the gluon-helicity distribution function at leading order. The 
helicity-dependent difference in hadron production is defined as:
\begin{eqnarray*}
\frac{d\Delta\sigma}{dp_{T}}&\equiv&
\frac{1}{2}\left[\frac{d\sigma^{++}}{dp_{T}}-
                 \frac{d\sigma^{+-}}{dp_{T}}\right],
\end{eqnarray*}
where the superscripts $++$ and $+-$ refer to same and opposite helicity 
combinations of the colliding protons~\cite{BunceAnnRev}. Factorization 
allows this to be written as a convolution of the long- and 
short-distance terms summed over all possible flavors for the partonic 
interaction $a+b\rightarrow c+X'$, where $c$ fragments into the 
detected hadron $h$:
\begin{eqnarray}
\frac{d\Delta\sigma}{dp_{T}}&=&
\sum\limits_{abc}\int dx_{a}dx_{b}dz_{c}~\Delta 
f_{a}(x_{a},\mu_{f})~\Delta f_{b}(x_{b},\mu_{f})\nonumber\\
&&\times\frac{d\Delta\hat{\sigma}^{ab\rightarrow cX'}}
{dp_{T}}(x_{a}P_{a},x_{b}P_{b},P^{h}/z_{c},\mu_{f},\mu_{f}',\mu_{r})\nonumber\\
&&\times D_{c}^{h}(z_{c},\mu_{f}'), 
\end{eqnarray}
where $\Delta f(x,\mu_f)$ are the polarized PDFs of the colliding 
partons carrying light-cone momentum fraction $x$ evaluated at 
factorization scale $\mu_{f}$. The fragmentation function of scattered 
parton $c$ into hadron $h$ with fraction $z_c$ of the scattered parton 
momentum is $D_{c}^{h}(z_{c},\mu_{f}')$ at fragmentation scale 
$\mu_{f}'$. The helicity-dependent difference in the cross section of 
the hard partonic scattering $a+b\rightarrow c+X'$ is denoted by 
$d\Delta\hat{\sigma}$ and is calculable in perturbative QCD. Cross 
section calculations to finite order in $\alpha_{s}$ have a dependence 
on factorization and renormalization scales $\mu_{f}$ and $\mu_{r}$.

Instead of directly measuring the helicity-dependent cross-section 
difference $d\Delta\sigma/dp_{T}$, we extract the double-longitudinal 
spin asymmetry defined as the ratio of the polarized to unpolarized 
cross sections $A_{LL}\equiv d\Delta\sigma/d\sigma$.  Here, $d\sigma$ 
is the helicity-averaged (unpolarized) cross section $d\sigma\equiv 
[d\sigma_{(++)}+d\sigma_{(+-)}]/2$. The ratio $d\Delta\sigma/d\sigma$ 
has smaller systematic uncertainties since some of the uncertainties 
cancel.

At $\sqrt{s}=62.4$ GeV, the production of final-state hadrons at 
midrapidity in a transverse momentum range $1.5 \leq p_T \leq 4.5$ 
GeV/$c$ is dominated by quark-gluon scattering~\cite{subprocess}. This 
makes the asymmetry reported here, $A_{LL}(p$+$p\rightarrow h^{\pm}+X)$, 
sensitive to the polarized gluon PDF $\Delta G(x)$ at leading order, and 
more sensitive to its sign than processes dominated by gluon-gluon 
scattering or which produce isospin-symmetric particles.  For example, 
preferential fragmentation of up quarks into positive pions and down 
quarks into negative pions, combined with the fact that the up quark 
helicity PDF is positive and the down quark helicity PDF is negative, 
would lead to an ordering of the asymmetries of pions (charged and 
neutral) directly sensitive to the sign of the gluon helicity PDF. 
Positive $\Delta G(x)$ would lead to $A^{\pi^+}_{LL} \geq A^{\pi^0}_{LL} 
\geq A^{\pi^-}_{LL}$, whereas a negative $\Delta G(x)$ would imply an 
opposite ordering.  These results can be combined with data from 
polarized collider and fixed target experiments in a global analysis to 
reduce uncertainties on the gluon helicity 
distribution~\cite{DSSV,DSSV_prd}.

%----------------------

\section{Experimental Setup}

This analysis uses the PHENIX central arm spectrometers. Each arm has an 
acceptance covering a pseudorapidity range $|\eta|<0.35$ and 
$\Delta\phi=\frac{\pi}{2}$ in azimuth~\cite{NIMPH,NIMPHCA}. The PHENIX 
central magnet creates an axial magnetic field with $\int B dl = 0.78$ 
T$\cdot$m at $\frac{\pi}{2}$ in this region.

Midrapidity charged hadrons are tracked in the drift chambers (DC), 
which are located outside the magnetic field with an inner radius of 
2.0~m and outer radius of 2.4~m. The tracks are reconstructed as nearly 
straight lines and yield the deflection in the axial magnetic field to 
determine the transverse momentum with a resolution 
$\frac{dp_{T}}{p_{T}} = 0.007\oplus 0.009~p_{T}$ (GeV/$c$) 
\cite{PHENIXmomresnew}.  The first term in the resolution is dominated 
by multiple scattering in material before the DC, while the second 
arises from the finite angular resolution of the DC. The momentum scale 
is set by requiring the proton mass reconstructed from the measured 
momentum and time of flight to match the known proton mass.

Track reconstruction also utilizes two layers of pad chambers (PC), 
which are multiwire proportional chambers with pad 
readout~\cite{NIMPHCA}. The first layer, PC1, is located after the DC, 
with an average radius of $2.49$ m. PC1 information is used in 
conjunction with the DC hits and vertex information to determine the 
polar angle for each charged track. The outermost layer PC3 is at an 
average radius of $4.98$~m and is used for charged track selection by 
matching PC3 hit positions with tracks projected using information from 
the DC and PC1 and the measured event vertex. Matching projected tracks 
with hit positions in PC3 also helps in rejecting decay backgrounds from 
primary hadrons.

Vertex and timing information is provided by two beam-beam counters 
(BBCs)~\cite{NIMPH} placed around the beam pipe. The BBCs are located 
1.44~m forward and backward of the nominal interaction point. Each BBC 
comprises an array of 64 phototubes fitted with $3$ cm long quartz 
radiators. The phototubes detect \v{C}erenkov radiation from charged 
particles traversing the quartz. The detectors cover a pseudorapidity 
range of $3.0 \leq |\eta| \leq 3.9$, and a full $\Delta\phi=2\pi$ in 
azimuth. A coincidence of hits from the two BBCs forms the minimum bias 
trigger, with timing information providing the location of the event 
vertex along the beam line with a few cm precision.

To eliminate the $e^{\pm}$ background due to photon conversion in 
material before the DC (primarily the beam pipe and DC entrance window), 
the analysis uses information from a ring imaging \v{C}erenkov detector 
(RICH)~\cite{NIMPHCA} located after the DC. The RICH uses CO$_{2}$ at 
atmospheric pressure as a radiator, with a momentum threshold of 
$17$~MeV/$c$ for $e^{\pm}$ and $4.7$~GeV/$c$ for charged pions. A RICH 
veto (ensuring no RICH hits) is used to reject $e^{\pm}$ and results in 
an upper transverse-momentum limit of $p_{T}<4.5$ GeV/$c$ for charged 
hadrons in the analysis.

Zero degree calorimeters (ZDC), which detect neutral particles near the 
beam pipe ($\theta < 2.5$ mrad) are used in conjunction with the BBC to 
estimate the systematic uncertainty on the relative luminosity for the 
asymmetry measurements~\cite{PI062}. The BBC and ZDC are also used to 
determine the integrated luminosity measurement~\cite{vernier1, vernier2}.

The stable spin direction of polarized protons in RHIC is vertical. The 
spin direction can be rotated into the longitudinal direction at the 
PHENIX interaction region using pairs of spin rotators. The 
polarizations of the beams at RHIC are measured every few hours by 
carbon target polarimeters~\cite{RHICpol}, which are normalized to an 
absolute measurement with a hydrogen jet target 
polarimeter~\cite{HJetpol}.

%----------------------

\section{Cross Section}

The results presented here are the first measurements of the cross 
section of inclusive charged-hadron production at midrapidity in the 
transverse momentum range $0.5 \leq p_T \leq 4.5$ GeV/$c$ from $p$+$p$ 
collisions at $\sqrt{s} = 62.4$ GeV.  The analysis techniques are 
similar to methods described in Ref.~\cite{PHRun2ch} and are briefly 
explained in Sect.~\ref{sec:crossana}.  The cross section results are 
presented and discussed in Sect.~\ref{sec:crossresult}.

%----------
\subsection{Cross Section Measurement}
\label{sec:crossana}

Approximately $2.14 \times 10^8$ BBC-triggered events corresponding to 
an integrated luminosity of $15.6 \ \rm nb^{-1}$ from 
polarization-averaged $p+p$ data taken in 2006 have been analyzed. We 
calculate the midrapidity charged-hadron production cross section using 
the following formula:
\begin{eqnarray}
\frac{E}{c} \frac{d^3\sigma}{dp^3} = 
\frac{\sigma_\textrm{\rm BBC}}{N_\textrm{\rm BBC}}
\frac{d^3 N(p_T)}{d\phi p_T dp_T dy}R_\textrm{smear}C_\textrm{\rm trig}
\frac{1}{E^\textrm{acc}_\textrm{eff}}, 
\end{eqnarray}
where $\sigma_\textrm{\rm BBC}$ is the $p+p$ cross section seen by the BBC 
as measured in Ref.~\cite{PI062}, $N_\textrm{\rm BBC}$ is the total number 
of BBC-triggered events analyzed, $R_\textrm{smear}$ is the correction 
factor for the smearing of track $p_T$ owing to the momentum resolution 
of the detectors as well as multiple scattering of the hadron tracks, 
$C_\textrm{\rm trig}$ is the correction factor for BBC trigger bias and 
$E^\textrm{acc}_\textrm{eff}$ is the combined correction factor for 
geometrical acceptance of the detectors and reconstruction efficiency.

\begin{figure}[htb]
\includegraphics[width=0.99\linewidth]{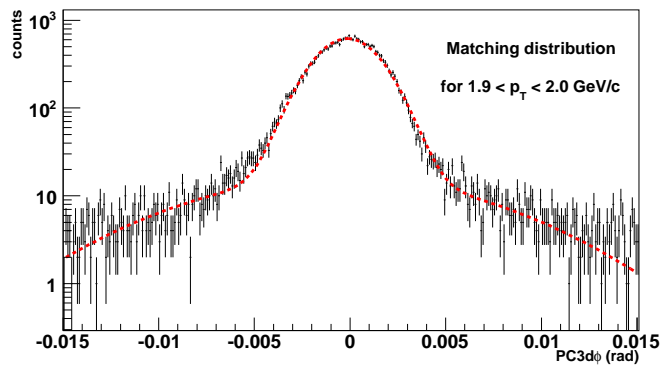}
\caption{(color online) 
Difference between the track-extrapolated and actual hit positions on 
PC3 in azimuth for a sample $p_T$ bin.  The histogram represents data; 
the dashed line represents a two-Gaussian fit.}
\label{fig:matching}
\end{figure}

The reconstructed charged tracks in the transverse momentum range $0.5 
\leq p_T \leq 4.5$ GeV/$c$ from events with vertices within $\pm \ 30$ 
cm of the nominal interaction point are matched to projected hit 
positions in PC3 in azimuthal ($\phi$) and beam direction ($z$). 
Distributions of the matching variables (difference between the 
projected track position and actual hits on PC3, termed PC3d$\phi$ and 
PC3d$z$) are fit with the combination of a signal Gaussian and a second 
Gaussian for background.  Figure~\ref{fig:matching} demonstrates the 
method for a sample $p_T$ bin. The width of the fit to the signal 
distribution is used to impose a simultaneous selection window of $2 
\sigma$ for both matching variables.

%%%%%%%%%%%%%%%%%%%%%%%%%%%%%%%%%%%%%%%%%%%%%  Fig_1

The background to these measurements comes from several sources. One 
source is soft electrons from the magnet pole faces, and another is 
decays in flight of $\pi^{\pm}$ and $K^{\pm}$. Not all of these 
electrons are rejected by the RICH cut. Off-vertex electrons and 
daughter particles with a perpendicular momentum kick from the decay are 
reconstructed as apparent high-$p_T$ tracks, but with wide Gaussian 
track matching distributions.  The background fraction in each of the 
$p_T$ bins is determined by using the distributions of the matching 
variables. Background fractions, which are 2--5\% for $p_T \leq 2.75$ 
GeV/$c$ and $\sim 30\%$ in the highest $p_T$ bin, are subsequently 
subtracted from hadron yields.

%%%%%%%%%%%%%%%%%%%%%%%%%%%%%%%%%%%%%%%%%%%%%  Fig_2
\begin{figure}[htb]
\includegraphics[width=0.99\linewidth]{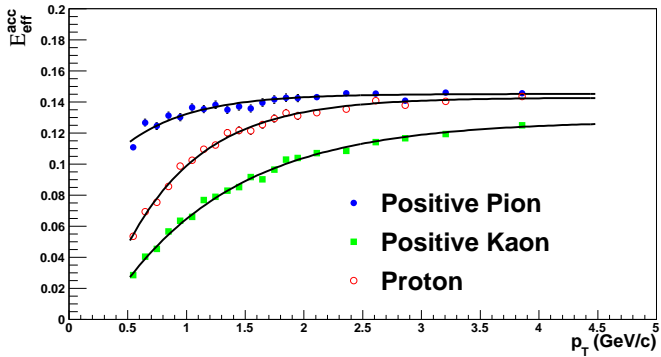}
\caption{(color online) Combined acceptance + detection efficiencies for 
separate hadron species.}
\label{fig:ideff}
\end{figure}

An additional source of background is the feed-down background produced 
by weak decays of mostly $\Lambda$ particles close to the event vertex 
with apparent momenta close to their true momenta and matching 
distributions peaked under the signal. Feed-down contributions to the 
detected protons and antiprotons from weak decays of $\Lambda$'s and 
heavier hyperons are estimated using input $\Lambda$ and 
$\overline{\Lambda}$ spectra from $p+p$ measurements at $\sqrt{s} = 63$ 
GeV at the ISR~\cite{ISRBSC, ISR_lambda} and at $\sqrt{s} = 62.4$ GeV at 
PHENIX with a {\sc geant}3~\cite{GEANT3}-based simulation of the PHENIX 
detector. The fractional contributions of the feed-down protons and 
antiprotons are independent of $p_T$ above $p_T = 2$~GeV/$c$ and are 
close to $7$ and $15\%$ respectively. Below $p_T = 2$ GeV/$c$ the 
fractions increase with decreasing $p_T$ and are roughly $25$ and $60 
\%$ for protons and antiprotons respectively at $p_T = 0.5$ GeV/$c$.

Background-subtracted yields are corrected for angular resolution of 
the DC and for smearing of the reconstructed momenta resulting from 
multiple scattering of tracks, which depends on hadron mass.  To 
account for the acceptance of the PHENIX detector system and the 
varying efficiency for different hadrons, single-particle 
Monte-Carlo simulations are performed and verified by comparing the 
live detector area between data and Monte Carlo, and appropriate 
correction factors are determined separately for each hadron 
species.

%%%%%%%%%%%%%%%%%%%%%%%%%%%%%%%%%%%%%%%%%%%%%%%%%%%%%%% Table_I
\begin{table}[htb]
\caption{Fit-function parameters for the efficiency curves for different 
hadron species. See text for details.}
	\begin{ruledtabular} \begin{tabular}{cccc}
	hadron & A & B & C\\
	\hline
	$\pi^+$ & -0.08 & -1.8  & 0.1453 \\
	$K^+$  & -0.17 & -0.97 & 0.1280 \\
	$p$	   & -0.21 & -1.54 & 0.1428 \\
	$\pi^-$ & -0.07 & -1.7  & 0.1449 \\
	$K^-$  & -0.17 & -1.01 & 0.1276 \\
	$p^-$  & -0.21 & -1.56 & 0.1424 \\
	\end{tabular} \end{ruledtabular} 
\label{table:effpar}
\end{table}

%%%%%%%%%%%%%%%%%%%%%%%%%%%%%%%%%%%%%%%%%%%%%%%%%%%%%%% Table_II
\begin{table}[htb]
\caption{Fit-function parameters for relative fractions of different 
species in the hadron mix. See text for details.}
	\begin{ruledtabular} \begin{tabular}{ccccc}
	hadron & A & B & C & D\\
	\hline
	$\pi^+$ & 1.02 & -2.39 & 0.57  & - \\
	$K^+$  & -0.53 & -2.39 & 0.20 & 0.009 \\
	$p$	   & -0.49 & -2.39 & 0.23 & -0.009 \\
	$\pi^-$ & 1.17 & -2.49	 & 0.61 & 0.012 \\
	$K^-$  & -0.61 & -2.49 & 0.20 & - \\
	$p^-$  & -0.56 & -2.49 & 0.18 & -0.012\\
	\end{tabular} \end{ruledtabular} 
\label{table:idfrac}
\end{table}

%%%%%%%%%%%%%%%%%%%%%%%%%%%%%%%%%%%%%%%%%%%%%%%%%%%%%%%%  Table_III
\begin{table}[htb]
\caption{Systematic uncertainties of cross section measurements from 
various sources.}
	\begin{ruledtabular} \begin{tabular}{cc}
	Source&Systematic Uncertainty\\
	\hline
	Acceptance \& efficiency correction&11--24\%\\
	$\sigma_{\rm BBC}$&11\%\\
	Trigger bias&2.5\%\\
	Monte Carlo/data scale factor&2\%\\
	PID fraction&1--5\%\\
	Background fraction&1--5\%\\
	Momentum smearing correction&1--2\%\\
	\end{tabular} \end{ruledtabular} 
\label{table:xsys}
\end{table}

%%%%%%%%%%%%%%%%%%%%%%%%%%%%%%%%%%%%%%%%%%%%%  Fig_3
\begin{figure*}[th]
\includegraphics[width=0.8\linewidth]{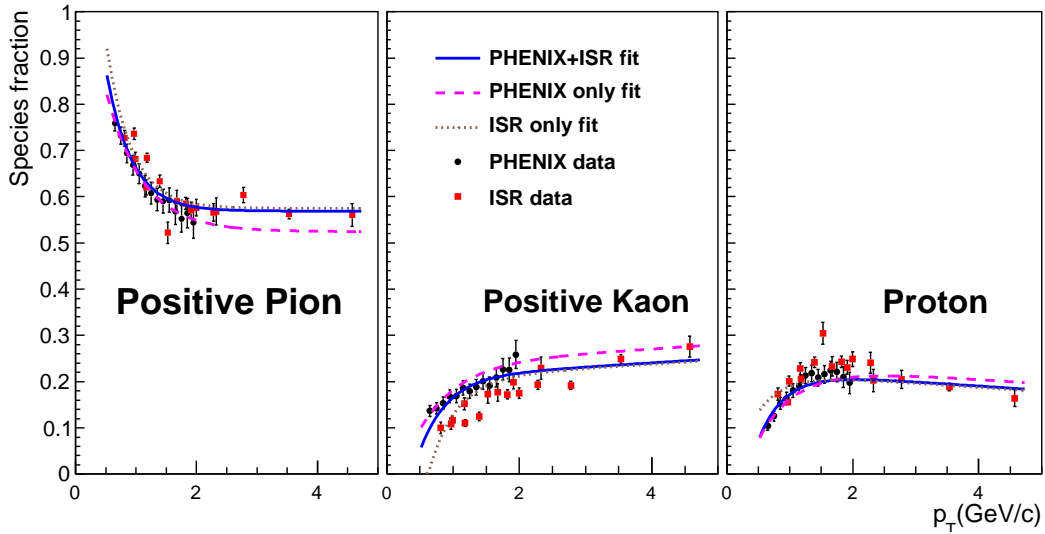}
\caption{(color online) Relative fraction of each species for positive 
hadrons.  The error bars on the data points are combined statistical and 
systematic uncertainties.  The solid line represents a fit to both the 
PHENIX and ISR data.  The dashed (dotted) line represents a fit to only 
the PHENIX (ISR) data.}
\label{fig:idfrac}
\end{figure*}
%-----------------------

Figure~\ref{fig:ideff} shows the efficiencies for the three positive 
hadron species.  The small efficiency for kaons is due to decays in 
flight. The large decrease in efficiency at low $p_T$ is due to the fact 
that the fixed pseudorapidity acceptance of the detector corresponds to a 
narrow range in rapidity for smaller ${p_T}/m$.  The efficiencies 
are parameterized ($Ae^{Bp_T} + C$) as a function of $p_T$. 
Table~\ref{table:effpar} shows the fit-function parameters. The fit values 
are used for the calculation of cross sections.  The species-dependent 
corrections are applied according to their production fraction 
(Fig.~\ref{fig:idfrac}) in the hadron mixture. The production fractions 
were determined from identified hadron spectra from 
PHENIX~\cite{PH62idch}, as well as earlier data from the 
ISR~\cite{ISRBSC}.  Dotted and dashed lines in Fig.~\ref{fig:ideff} 
represent fits with individual data sets.  Parameterized fits of the form 
$Ae^{Bp_T} + C$ (for positive pions and negative kaons) and $Ae^{Bp_T} + C
+ Dp_T$ (for all other species) are performed under the constraint that 
the sum of the relative fractions is $1$. 
Table~\ref{table:idfrac} gives the fit-function parameters.  
Relative fractions from fit values are used to 
apply the corrections.  The corrected yields are scaled by the BBC trigger 
bias as described in Ref.~\cite{PI062}.

The integrated luminosity $\mathcal{L}_{int} = 
\frac{N_{\rm BBC}}{\sigma_{\rm BBC}}$, required for normalization of the 
invariant cross section, is calculated using the count of BBC-triggered 
events and the BBC normalization parameter $\sigma_{\rm BBC}$. The parameter 
$\sigma_{\rm BBC}$ is the $p+p$ cross section seen by the BBC and is 
measured by the Van der Meer (Vernier) scan technique~\cite{vernier2}. 
The quantity $\sigma_{\rm BBC} = 13.7 \pm 1.5 \ \rm^{\rm syst}$ mb for the 
relevant data set has been measured for previous PHENIX results and was 
discussed in detail in Ref.~\cite{PI062}.

Table~\ref{table:xsys} shows the 
systematic uncertainties of cross section measurements from
various sources.
The largest contribution (11--24\%) to the $p_T$-dependent systematic 
uncertainty comes from the correction for the acceptance and detection 
efficiencies.  The uncertainty on the cross section due to this 
correction factor is determined by varying the selection parameters in 
the MC simulations.  The trigger bias introduces a $2.5\%$ uncertainty 
in the overall normalization, in addition to the 11\% uncertainty on 
$\sigma_{\rm BBC}$.  Determination of the background fraction and the 
production fraction of separate hadron species each introduces a
1--5 \% $p_T$-dependent systematic uncertainty. Uncertainties from other 
sources, for example the correction for momentum resolution, the 
correction for the active area of the detector in experiment and Monte 
Carlo, are $\sim$1--2\%.

%----------
\subsection{Cross Section Results}
\label{sec:crossresult}

\begin{figure*}[th]
\includegraphics[width=0.46\linewidth]{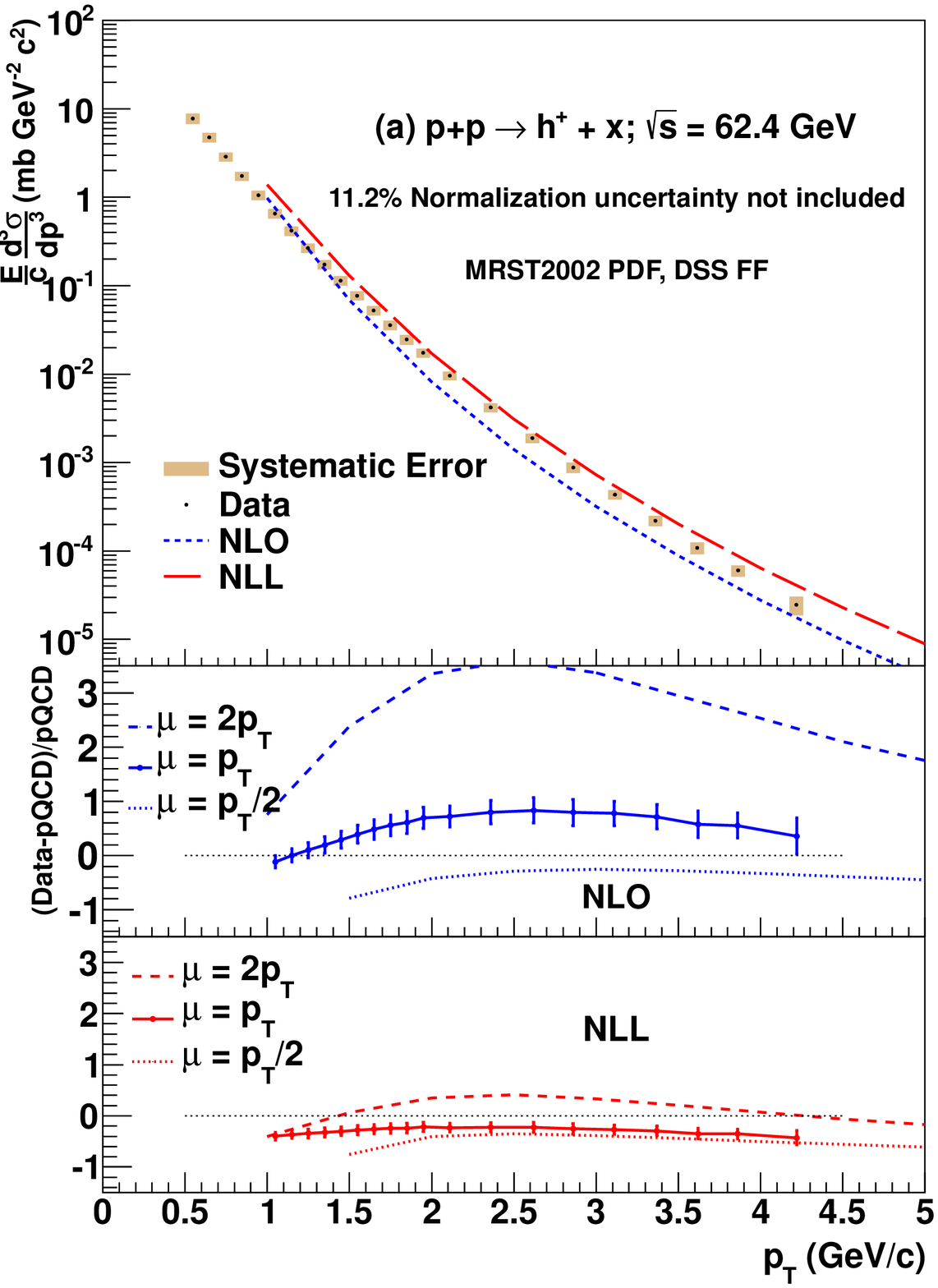} 
\includegraphics[width=0.46\linewidth]{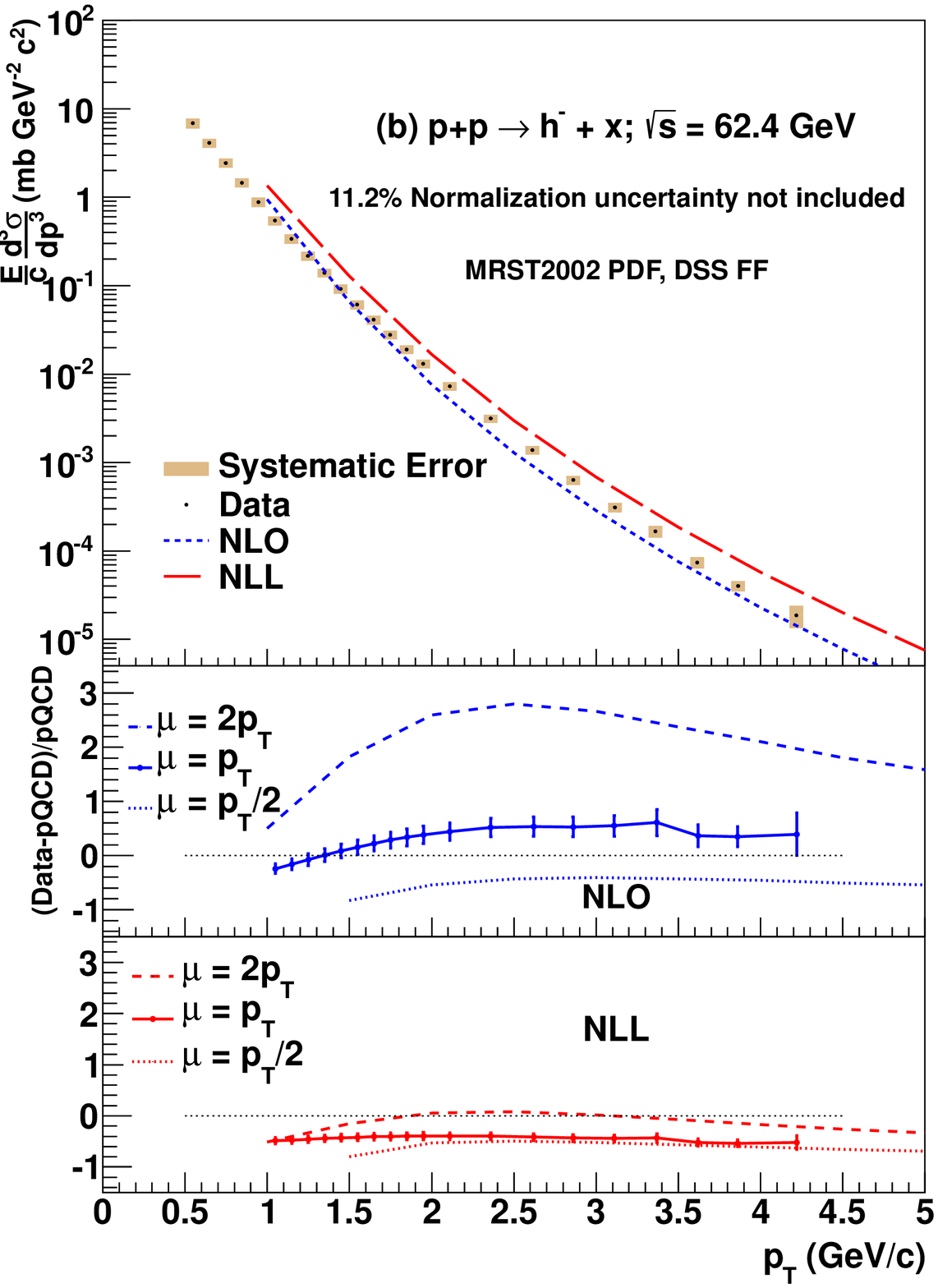}
\caption{(color online) 
Cross section of inclusive-charged-hadron production at midrapidity in 
$p+p$ at 62.4 GeV for (a) positive and (b) negative hadrons.  NLO and 
NLL theoretical predictions~\cite{pQCDXsec} at midrapidity, using 
MRST2002 parton distribution functions~\cite{MRST2002} and DSS 
fragmentation functions~\cite{DSSfrag2}, at factorization, 
renormalization, and fragmentation scale $\mu = p_T$ are shown as 
curves.  The lower panels show the scale dependence of the NLO and NLL 
results.
}
\label{fig:xsec} 
\end{figure*}

Figure~\ref{fig:xsec} and Table~\ref{table:feed} show the inclusive 
charged-hadron cross sections from $p+p$ collisions at $\sqrt{s} = 
62.4$ GeV as a function of $p_T$.  A combined $p_T$-independent 
normalization uncertainty of $11.2\%$ (uncertainties in the 
measurements of $\sigma_{\rm BBC}$ and BBC trigger bias) is not shown.

In the overlapping $p_T$ range, the results were found to be 
consistent with the species-combined cross sections from identified 
results at PHENIX~\cite{PH62idch} as well as ISR results at 
$\sqrt{s} = 63$~GeV~\cite{ISRBSC}.  On the upper panels of both 
plots in Fig.~\ref{fig:xsec} the cross sections are compared to NLO 
and NLL calculations at a factorization, renormalization and 
fragmentation scale of $\mu = p_T$~\cite{pQCDXsec}.  The 
calculations were performed using Martin-Roberts-Stirling-Thorne 
(MRST2002) PDFs~\cite{MRST2002} and deFlorian-Sassot-Stratmann (DSS) 
fragmentation functions~\cite{DSSfrag2}. The NLO predictions have 
been shown to describe midrapidity cross section results for neutral 
pions~\cite{PHENIX1, PHENIX2} and charged hadrons~\cite{PHRun2ch} at 
$\sqrt{s} = 200$ GeV within $\sim 20\%$ for a scale choice of $\mu 
= p_T$. For the present results at $\sqrt{s} = 62.4$ GeV, however, 
the NLO calculations underpredict the data by as much as $\sim 80\%$ 
in the case of positive hadrons and $\sim 60\%$ in the case of 
negative hadrons for the same scale choice. The lower two panels in 
the Fig.~\ref{fig:xsec} plots show the dependence of the theoretical 
calculations on the choice of factorization, renormalization and 
fragmentation scale ($\mu$) for three different values 
($p_T,~{p_T}/2, and 2p_T$). The inclusion of higher-order terms in 
the NLL calculations leads to a considerably smaller scale dependence.

%%%%%%%%%%%%%%%%%%%%%%%%%%%%%%%%%%%%%%%%%%%%%%%%%%%%  Fig_4

These new data at an energy intermediate to typical fixed-target and 
collider energies are timely, as the details of how to work with 
resummation techniques in different kinematic regimes are currently 
being explored by the theoretical community (see for 
example~\cite{Resum, Resum1, Resum2}). Comparison of the present 
results to the calculations at NLO with and without NLL terms included 
indicates that in the measured kinematic range, NLL terms make relevant 
contributions to the cross section. However, the tendency of the NLL 
calculations to overpredict the data, by as much as $\sim 40\%$ in the 
case of positive hadrons and $\sim 50\%$ in the case of negative 
hadrons for a scale choice of $p_T$, may indicate that there are terms 
in the full next-to-next-to-leading-order (NNLO) expansion that are of 
comparable magnitude and opposite sign to those in the NLL calculation. 
These measurements corroborate similar indications from neutral pion 
cross section results~\cite{PI062} and identified hadron cross section 
results~\cite{PH62idch} at PHENIX. The present measurements can also be 
useful in a future determination of inclusive charged-hadron 
fragmentation functions, as progress in pQCD has allowed inclusion of 
$p+p$ cross section measurements and semi-inclusive deep-inelastic 
lepton-nucleon scattering data in FF parameterizations along with the 
traditionally used $e^+e^-$ data since 
2007~\cite{DSSfrag,DSSfrag2,Etafrag}.

\begingroup \squeezetable
%%%%%%%%%%%%%%%%%%%%%%%%%%%%%%%%%%%%%%%%%%%%%%%%%%%%%%%%  Table_IV
\begin{table*}[bth]
\caption{Cross section of midrapidity charged-hadron production from 
$p+p$ collisions at $\sqrt{s} = 62.4$~GeV as a function of $p_T$.  The 
errors represent the statistical (first) and systematic uncertainties.  
The data are corrected for the contribution of feed-down protons and 
antiprotons.  A normalization uncertainty of $11.2\%$ is not included.}
\begin{ruledtabular} \begin{tabular}{ccc}
$p_T$& $h^+$ & $h^-$\\
(GeV/$c$)& mb GeV$^{-2}$ $c^2$ & mb GeV$^{-2}$ $c^2$\\
\hline
$0.55$&$7.80 \pm 1.2\times 10^{-03} \pm 1.1$&$6.87 \pm 1.1\times 10^{-03} \pm 9.3\times 10^{-01}$\\
$0.65$&$4.78 \pm 8.7\times 10^{-04} \pm 5.9\times 10^{-01}$&$4.10 \pm 8.0\times 10^{-04} \pm 5.0\times 10^{-01}$\\
$0.75$&$2.87 \pm 6.3\times 10^{-04} \pm 3.5\times 10^{-01}$&$2.45 \pm 5.7\times 10^{-04} \pm 2.9\times 10^{-01}$\\
$0.85$&$1.73 \pm 4.6\times 10^{-04} \pm 2.1\times 10^{-01}$&$1.46 \pm 4.2\times 10^{-04} \pm 1.7\times 10^{-01}$\\
$0.95$&$1.06 \pm 3.4\times 10^{-04} \pm 1.3\times 10^{-01}$&$8.83\times 10^{01} \pm 3.1\times 10^{-04} \pm 1.1\times 10^{-01}$\\
$1.05$&$6.55\times 10^{-01} \pm 2.5\times 10^{-04} \pm 8.2\times 10^{-02}$&$5.43\times 10^{-01} \pm 2.3\times 10^{-04} \pm 6.6\times 10^{-02}$\\
$1.15$&$4.18\times 10^{-01} \pm 1.9\times 10^{-04} \pm 5.2\times 10^{-02}$&$3.40\times 10^{-01} \pm 1.7\times 10^{-04} \pm 4.1\times 10^{-02}$\\
$1.25$&$2.67\times 10^{-01} \pm 1.5\times 10^{-04} \pm 3.4\times 10^{-02}$&$2.16\times 10^{-01} \pm 1.3\times 10^{-04} \pm 2.6\times 10^{-02}$\\
$1.35$&$1.73\times 10^{-01} \pm 1.2\times 10^{-04} \pm 2.2\times 10^{-02}$&$1.41\times 10^{-01} \pm 1.0\times 10^{-04} \pm 1.6\times 10^{-02}$\\
$1.45$&$1.14\times 10^{-01} \pm 9.0\times 10^{-05} \pm 1.4\times 10^{-02}$&$9.21\times 10^{-02} \pm 8.0\times 10^{-05} \pm 1.1\times 10^{-02}$\\
$1.55$&$7.73\times 10^{-02} \pm 7.2\times 10^{-05} \pm 9.3\times 10^{-03}$&$6.11\times 10^{-02} \pm 6.3\times 10^{-05} \pm 7.2\times 10^{-03}$\\
$1.65$&$5.25\times 10^{-02} \pm 5.7\times 10^{-05} \pm 6.4\times 10^{-03}$&$4.11\times 10^{-02} \pm 5.0\times 10^{-05} \pm 4.7\times 10^{-03}$\\
$1.75$&$3.59\times 10^{-02} \pm 4.6\times 10^{-05} \pm 4.4\times 10^{-03}$&$2.79\times 10^{-02} \pm 4.0\times 10^{-05} \pm 3.3\times 10^{-03}$\\
$1.85$&$2.46\times 10^{-02} \pm 3.7\times 10^{-05} \pm 3.1\times 10^{-03}$&$1.90\times 10^{-02} \pm 3.2\times 10^{-05} \pm 2.2\times 10^{-03}$\\
$1.95$&$1.74\times 10^{-02} \pm 3.0\times 10^{-05} \pm 2.0\times 10^{-03}$&$1.31\times 10^{-02} \pm 2.6\times 10^{-05} \pm 1.5\times 10^{-03}$\\
$2.11$&$9.61\times 10^{-03} \pm 1.4\times 10^{-05} \pm 1.1\times 10^{-03}$&$7.32\times 10^{-03} \pm 1.2\times 10^{-05} \pm 8.3\times 10^{-04}$\\
$2.36$&$4.19\times 10^{-03} \pm 8.5\times 10^{-06} \pm 5.0\times 10^{-04}$&$3.14\times 10^{-03} \pm 7.3\times 10^{-06} \pm 3.5\times 10^{-04}$\\
$2.61$&$1.89\times 10^{-03} \pm 5.5\times 10^{-06} \pm 2.4\times 10^{-04}$&$1.38\times 10^{-03} \pm 4.6\times 10^{-06} \pm 1.6\times 10^{-04}$\\
$2.86$&$8.80\times 10^{-04} \pm 3.6\times 10^{-06} \pm 1.2\times 10^{-04}$&$6.36\times 10^{-04} \pm 3.0\times 10^{-06} \pm 7.5\times 10^{-05}$\\
$3.11$&$4.33\times 10^{-04} \pm 2.4\times 10^{-06} \pm 5.4\times 10^{-05}$&$3.12\times 10^{-04} \pm 2.0\times 10^{-06} \pm 3.8\times 10^{-05}$\\
$3.37$&$2.20\times 10^{-04} \pm 1.6\times 10^{-06} \pm 2.9\times 10^{-05}$&$1.67\times 10^{-04} \pm 1.4\times 10^{-06} \pm 2.5\times 10^{-05}$\\
$3.62$&$1.09\times 10^{-04} \pm 1.1\times 10^{-06} \pm 1.7\times 10^{-05}$&$7.41\times 10^{-05} \pm 9.1\times 10^{-07} \pm 1.1\times 10^{-05}$\\
$3.87$&$6.03\times 10^{-05} \pm 8.1\times 10^{-07} \pm 9.0\times 10^{-06}$&$4.02\times 10^{-05} \pm 6.4\times 10^{-07} \pm 5.5\times 10^{-06}$\\
$4.22$&$2.46\times 10^{-05} \pm 3.5\times 10^{-07} \pm 6.0\times 10^{-06}$&$1.88\times 10^{-05} \pm 2.9\times 10^{-07} \pm 5.3\times 10^{-06}$\\
\end{tabular} \end{ruledtabular}
\label{table:feed}
\end{table*}
\endgroup

%---------------------

\section{Double-Helicity Asymmetry}

We measured the double-helicity asymmetries, $A_{LL}$, of inclusive 
positive- and negative-hadron production in the transverse-momentum 
range of $0.5 \leq p_T \leq 4.5$ GeV/$c$ at midrapidity from 
longitudinally polarized $p+p$ collisions at $\sqrt{s} = 62.4$ GeV.

%----------
\subsection{$A_{LL}$ Measurement}
\label{sec:ALLana}

The double-helicity asymmetry of charged hadrons is defined as the 
relative difference between hadron-production cross sections from 
collisions of the same- and opposite-helicity state protons. 
Experimentally, the asymmetry is measured as

\begin{equation}
A_{LL} = \frac{1}{P_B \cdot P_Y} \frac{N_{++} 
- R \cdot N_{+-}}{N_{++} + R \cdot N_{+-}},
\end{equation}
where $P_B$, $P_Y$ are polarizations of the two colliding beams in RHIC 
(termed `Blue' and `Yellow'), $N_{++}$, $N_{+-}$ are the midrapidity 
hadron yields from collisions of the same- and opposite-helicity 
protons, and relative luminosity $R = \frac{L_{++}}{L_{+-}}$ is the 
ratio of luminosity of the same-helicity collisions to that of 
opposite-helicity collisions.

For the 2006 $p+p$ data set at $\sqrt{s} = 62.4$ GeV, the 
luminosity-weighted average beam polarizations for both beams are 
measured to be $\left\langle P \right\rangle = 0.48$, and the average 
magnitude of the product of polarization of the two beams is 
$\left\langle P_B \cdot P_Y \right\rangle = 0.23$ with a relative 
uncertainty of $13.9\%$. The colliding proton bunches at RHIC are 
assigned preset spin patterns repeated every four crossings. For 
consecutive fills with $120$ bunches in the RHIC ring four such 
different spin patterns are alternated in order to reduce false 
asymmetries and systematic effects of possible correlations between the 
detector response and the RHIC bunch structures.

Hadron counts are obtained under similar criteria as described for the 
cross section measurements from the reconstructed tracks in 
Sect.~\ref{sec:crossana}. Approximately $1.63 \times 10^8$ BBC-triggered 
events with longitudinal beam polarization, corresponding to an 
integrated luminosity of $11.9 \ \rm nb^{-1}$, were analyzed for the 
asymmetry measurements. Luminosities for the same- and opposite-helicity 
events were obtained from crossing-by-crossing information of BBC 
trigger counts. The systematic uncertainty of the relative luminosity 
$R$, determined by comparing BBC-triggered events with events triggered 
by the ZDCs, was found to be $1.4 \times 10^{-3}$.

The asymmetry is determined on a fill-by-fill basis. The asymmetry of 
backgrounds from decays in flight, selected from the tail ends of the 
distribution of the matching variables (in beam direction $z$ and 
azimuthal angle $\phi$), is measured. The background asymmetry 
$A^\textrm{bkg}_{LL}$ and background fraction are used to calculate the 
signal asymmetry and its statistical uncertainty. The feed-down from 
decays of $\Lambda$'s and heavier hyperons, however, cannot be separated 
from hadron yields. The backgrounds from feed-down protons and 
antiprotons are a small contribution (1--2.5\%) to the total hadron 
yields and are not corrected for since the asymmetries of these 
backgrounds are unknown.

%----------
\subsection{$A_{LL}$ Results}
\label{sec:ALLresult}

\begin{figure*}[tbh]
\includegraphics[width=0.49\linewidth]{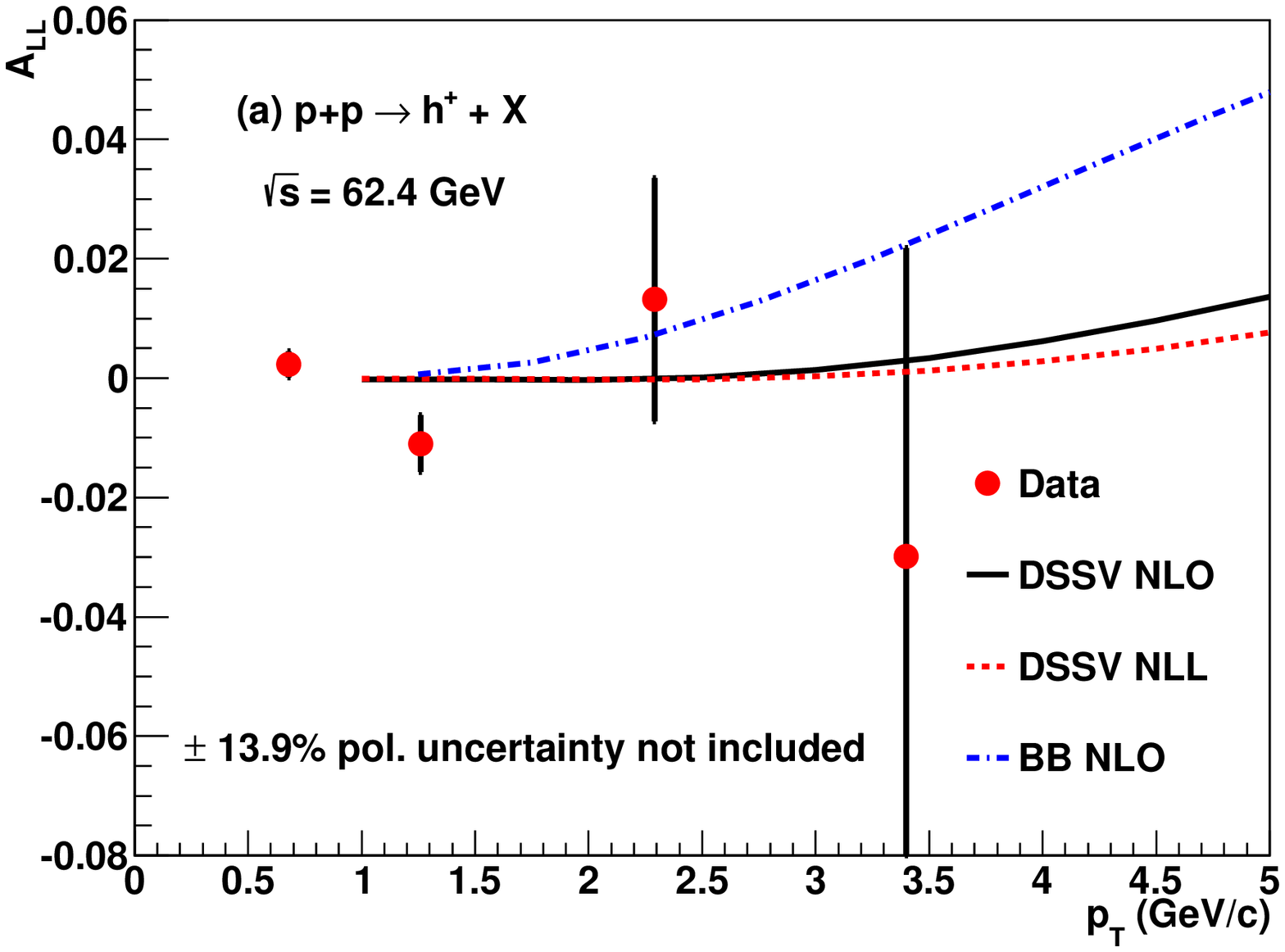}
\includegraphics[width=0.49\linewidth]{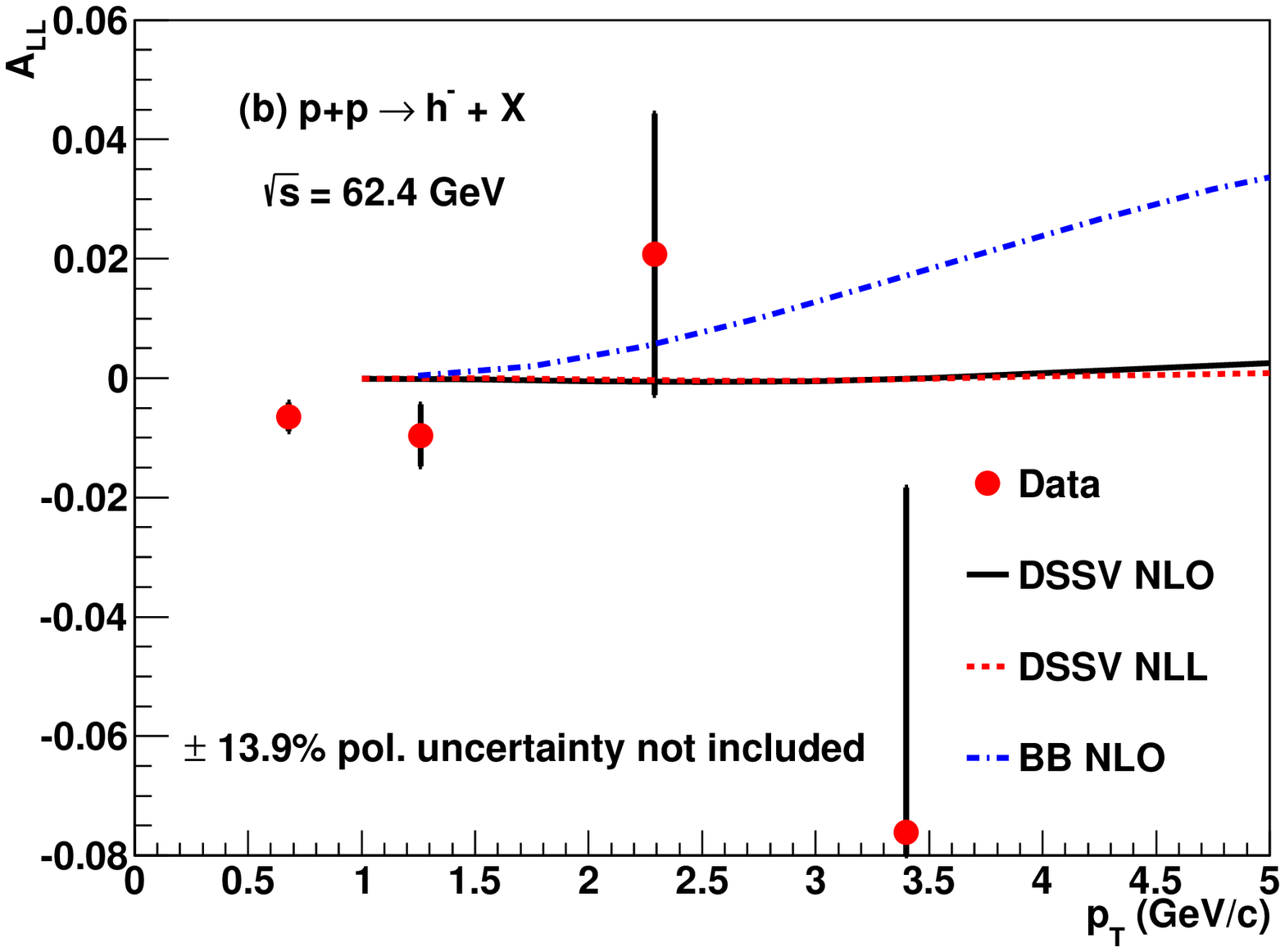}
\caption{(color online) 
Double-helicity asymmetry ($A_{LL}$) of (a) positive and (b) 
negative charged-hadron production from polarized $p$+$p$ collisions 
at $\sqrt{s}=62.4$ GeV.  The results are compared to NLO and NLL pQCD 
predictions using several parameterizations of the helicity PDFs 
(see text for details). 
}
\label{fig:all} 
\end{figure*}

Figure~\ref{fig:all} and Table~\ref{table:a_ll} show the 
$p_T$-dependence of the measured double-helicity asymmetries for 
inclusive-charged-hadron production at midrapidity in polarized 
$p$+$p$ collisions at $\sqrt{s} = 62.4$ GeV.  The asymmetries are 
compared to NLO pQCD predictions~\cite{NLOALL} based on several 
different parameterizations of polarized PDFs at scale $\mu = p_T$.  

The curves in Fig.~\ref{fig:all} labeled ``DSSV NLO" and ``DSSV 
NLL" refer to deFlorian-Sassot-Stratmann-Vogelsang (DSSV) 
parameterizations of the helicity PDFs~\cite{DSSV} and curves 
labeled ``BB NLO" refer to Bl\"{u}mlein-B\"{o}ttcher (BB) 
parameterizations of the helicity PDFs~\cite{BB}.  Both DSSV and BB 
calculations use DSS fragmentation functions~\cite{DSSfrag}.  
DSSV calculations use MRST2002 unpolarized PDFs~\cite{MRST2002}; 
BB calculations use those from the 
coordinated-theoretical-experimental-project-on-QCD-6~\cite{CTEQ6}.  
Polarized PDFs use fits to the pDIS data to extract 
parameters of the functional forms of PDFs.  DSSV~\cite{DSSV} 
parameterizations use RHIC data along with the available pDIS data 
to constrain the polarized PDFs.  The asymmetries are also compared 
to the NLL estimations with the DSSV PDFs~\cite{pQCDXsec}.

%%%%%%%%%%%%%%%%%%%%%%%%%%%%%%%%%%%%%%%%%%%%%%%%%%%%%%%%  Fig_5

%%%%%%%%%%%%%%%%%%%%%%%%%%%%%%%%%%%%%%%%%%%%%%%%%%%%%%%%  Table_V
\begin{table*}[tbh]
\caption{The double-helicity asymmetries and the statistical 
uncertainties as a function of $p_T$ for positive and negative inclusive 
charged hadrons from $p+p$ collisions at $\sqrt{s} = 62.4$~GeV. The 
fractional contribution to the yields from weak-decay feed-down to 
protons and antiprotons is shown; no correction to the asymmetries has 
been made for these contributions.}
\begin{ruledtabular} \begin{tabular}{ccccc}
    &  & estimated &  &  estimated \\
  $p_T$ (GeV/$c$)
& $A_{LL} \pm \delta A_{LL}$($h^+$) 
& feed-down fraction($h^+$) 
& $A_{LL} \pm \delta A_{LL}$($h^-$)
& feed-down fraction($h^-$)\\
\hline
$0.68$&$0.0023 \pm 0.0022$&$0.022$&$-0.0065 \pm 0.0024$&$0.021 $\\
$1.26$&$-0.01096 \pm 0.0048$&$0.016$&$-0.0096 \pm 0.0052$&$0.025 $\\
$2.29$&$0.0132 \pm 0.0204$&$0.012$&$0.0208 \pm 0.0236$&$0.021 $\\
$3.40$&$-0.0299 \pm 0.0517$&$0.011$&$-0.0761 \pm 0.0578$&$0.018 $\\
\end{tabular} \end{ruledtabular}
\label{table:a_ll}
\end{table*}

For the purpose of comparison with the experimental results, pQCD 
calculations were obtained for separate hadron species (pions, kaons and 
(anti)protons) and were combined using the particle fraction in the 
hadron mixture (Fig.~\ref{fig:idfrac}) and corresponding efficiency 
factor (Fig.~\ref{fig:ideff}). The measured asymmetries are small and 
consistent with zero. The results are also consistent with the 
predictions from the recent parameterizations within statistical 
limitations. The comparisons corroborate previous PHENIX 
measurements~\cite{PI062, PI0200} that disfavor very large gluon 
polarization. The presented asymmetry measurements probe a range of 
approximately $0.05 \leq x_{gluon} \leq 0.2$~\cite{Resum} of the 
interacting gluons.

%----------------------

\section{Summary and Conclusions}

Cross sections and double-helicity asymmetries for the midrapidity 
production of positive and negative inclusive charged hadrons at 
$\sqrt{s}=62.4$ GeV are measured as a function of transverse 
momentum.  The comparison with pQCD calculations shows that the NLO 
estimations are consistent with cross section results within a large 
scale uncertainty.  NLL calculations with their reduced scale dependence 
are also consistent with the data, indicating that the threshold 
resummation of logarithmic terms is relevant in the kinematic region 
measured; however, the overprediction of the data by up to $\sim 50\%$ 
if the NLL terms are included suggests that contributions from NNLO 
terms may also be important. This corroborates other recent results from 
PHENIX~\cite{PI062, PH62idch} with similar indications. The asymmetry 
results are the first measurements for charged hadron production in 
polarized $p+p$ collisions at $\sqrt{s}=62.4$ GeV and are consistent 
with the asymmetries found using several other probes at different 
collision energies at RHIC~\cite{PI062, PI0200, PHENIX2, STAR1, STAR2, 
STAR3}. Experimental measurements of a variety of processes covering a 
broad kinematic range are essential to advancing our understanding of 
QCD in hadronic interactions and nucleon structure, and the present 
measurements contribute towards that end.

%---------------
\begin{acknowledgments}

%\section{Acknowledgements}   % Run-6 long from for PRC, PLB, etc.

We thank the staff of the Collider-Accelerator and Physics
Departments at Brookhaven National Laboratory and the staff of
the other PHENIX participating institutions for their vital
contributions.  
We also thank Daniel de Florian, Werner Vogelsang and 
Federico Wagner for providing calculations, as well as 
Ted Rogers and Werner Vogelsang for valuable, in-depth discussions.
We acknowledge support from the 
Office of Nuclear Physics in the
Office of Science of the Department of Energy,
the National Science Foundation, 
a sponsored research grant from Renaissance Technologies LLC, 
Abilene Christian University Research Council, 
Research Foundation of SUNY, 
and Dean of the College of Arts and Sciences, Vanderbilt University 
(U.S.A),
Ministry of Education, Culture, Sports, Science, and Technology
and the Japan Society for the Promotion of Science (Japan),
Conselho Nacional de Desenvolvimento Cient\'{\i}fico e
Tecnol{\'o}gico and Funda\c c{\~a}o de Amparo {\`a} Pesquisa do
Estado de S{\~a}o Paulo (Brazil),
Natural Science Foundation of China (P.~R.~China),
Ministry of Education, Youth and Sports (Czech Republic),
Centre National de la Recherche Scientifique, Commissariat
{\`a} l'{\'E}nergie Atomique, and Institut National de Physique
Nucl{\'e}aire et de Physique des Particules (France),
Ministry of Industry, Science and Tekhnologies,
Bundesministerium f\"ur Bildung und Forschung, Deutscher
Akademischer Austausch Dienst, and Alexander von Humboldt Stiftung (Germany),
Hungarian National Science Fund, OTKA (Hungary), 
Department of Atomic Energy and Department of Science and Technology (India),
Israel Science Foundation (Israel), 
National Research Foundation and WCU program of the 
Ministry Education Science and Technology (Korea),
Ministry of Education and Science, Russian Academy of Sciences,
Federal Agency of Atomic Energy (Russia),
VR and the Wallenberg Foundation (Sweden), 
the U.S. Civilian Research and Development Foundation for the
Independent States of the Former Soviet Union, 
the US-Hungarian Fulbright Foundation for Educational Exchange,
and the US-Israel Binational Science Foundation.

\end{acknowledgments}
%---------------

%===========================
%\bibliography{ppg134x0}

\begin{thebibliography}{57}
\expandafter\ifx\csname natexlab\endcsname\relax\def\natexlab#1{#1}\fi
\expandafter\ifx\csname bibnamefont\endcsname\relax
  \def\bibnamefont#1{#1}\fi
\expandafter\ifx\csname bibfnamefont\endcsname\relax
  \def\bibfnamefont#1{#1}\fi
\expandafter\ifx\csname citenamefont\endcsname\relax
  \def\citenamefont#1{#1}\fi
\expandafter\ifx\csname url\endcsname\relax
  \def\url#1{\texttt{#1}}\fi
\expandafter\ifx\csname urlprefix\endcsname\relax\def\urlprefix{URL }\fi
\providecommand{\bibinfo}[2]{#2}
\providecommand{\eprint}[2][]{\url{#2}}

\bibitem[{\citenamefont{Ellis et~al.}(1978)\citenamefont{Ellis, Georgi,
  Machacek, Politzer, and Ross}}]{FactPropose}
\bibinfo{author}{\bibfnamefont{R.~K.} \bibnamefont{Ellis}},
  \bibinfo{author}{\bibfnamefont{H.}~\bibnamefont{Georgi}},
  \bibinfo{author}{\bibfnamefont{M.}~\bibnamefont{Machacek}},
  \bibinfo{author}{\bibfnamefont{H.~D.} \bibnamefont{Politzer}},
  \bibnamefont{and} \bibinfo{author}{\bibfnamefont{G.~G.} \bibnamefont{Ross}},
  \bibinfo{journal}{Phys. Lett. B} \textbf{\bibinfo{volume}{78}},
  \bibinfo{pages}{281} (\bibinfo{year}{1978}).

\bibitem[{\citenamefont{Collins et~al.}(1988)\citenamefont{Collins, Soper, and
  Sterman}}]{FactProof}
\bibinfo{author}{\bibfnamefont{J.~C.} \bibnamefont{Collins}},
  \bibinfo{author}{\bibfnamefont{D.~E.} \bibnamefont{Soper}}, \bibnamefont{and}
  \bibinfo{author}{\bibfnamefont{G.}~\bibnamefont{Sterman}},
  \bibinfo{journal}{Nucl. Phys. B} \textbf{\bibinfo{volume}{308}},
  \bibinfo{pages}{833} (\bibinfo{year}{1988}).

\bibitem[{\citenamefont{de~Florian and Vogelsang}(2005)}]{Resum1}
\bibinfo{author}{\bibfnamefont{D.}~\bibnamefont{de~Florian}} \bibnamefont{and}
  \bibinfo{author}{\bibfnamefont{W.}~\bibnamefont{Vogelsang}},
  \bibinfo{journal}{Phys. Rev. D.} \textbf{\bibinfo{volume}{71}},
  \bibinfo{pages}{114004} (\bibinfo{year}{2005}).

\bibitem[{\citenamefont{Adare et~al.}(2009{\natexlab{a}})}]{PI0200}
\bibinfo{author}{\bibfnamefont{A.}~\bibnamefont{Adare}} \bibnamefont{et~al.}
  (\bibinfo{collaboration}{PHENIX Collaboration}), \bibinfo{journal}{Phys. Rev.
  Lett.} \textbf{\bibinfo{volume}{103}}, \bibinfo{pages}{082002}
  (\bibinfo{year}{2009}{\natexlab{a}}).

\bibitem[{\citenamefont{Abelev et~al.}(2009)}]{STAR3}
\bibinfo{author}{\bibfnamefont{B.~I.} \bibnamefont{Abelev}}
  \bibnamefont{et~al.} (\bibinfo{collaboration}{STAR Collaboration}),
  \bibinfo{journal}{Phys. Rev. D} \textbf{\bibinfo{volume}{80}},
  \bibinfo{pages}{111108(R)} (\bibinfo{year}{2009}).

\bibitem[{\citenamefont{Adare et~al.}(2011{\natexlab{a}})}]{PHENIXJet}
\bibinfo{author}{\bibfnamefont{A.}~\bibnamefont{Adare}} \bibnamefont{et~al.}
  (\bibinfo{collaboration}{PHENIX Collaboration}), \bibinfo{journal}{Phys. Rev.
  D} \textbf{\bibinfo{volume}{84}}, \bibinfo{pages}{012006}
  (\bibinfo{year}{2011}{\natexlab{a}}).

\bibitem[{\citenamefont{Adler et~al.}(2006)}]{PHENIX3}
\bibinfo{author}{\bibfnamefont{S.~S.} \bibnamefont{Adler}} \bibnamefont{et~al.}
  (\bibinfo{collaboration}{PHENIX Collaboration}), \bibinfo{journal}{Phys. Rev.
  Lett.} \textbf{\bibinfo{volume}{97}}, \bibinfo{pages}{052301}
  (\bibinfo{year}{2006}).

\bibitem[{\citenamefont{Abelev et~al.}(2006)}]{STAR1}
\bibinfo{author}{\bibfnamefont{B.~I.} \bibnamefont{Abelev}}
  \bibnamefont{et~al.} (\bibinfo{collaboration}{STAR Collaboration}),
  \bibinfo{journal}{Phys. Rev. Lett.} \textbf{\bibinfo{volume}{97}},
  \bibinfo{pages}{252001} (\bibinfo{year}{2006}).

\bibitem[{\citenamefont{Adler et~al.}(2007)}]{PHENIX4}
\bibinfo{author}{\bibfnamefont{S.~S.} \bibnamefont{Adler}} \bibnamefont{et~al.}
  (\bibinfo{collaboration}{PHENIX Collaboration}), \bibinfo{journal}{Phys. Rev.
  Lett.} \textbf{\bibinfo{volume}{98}}, \bibinfo{pages}{012002}
  (\bibinfo{year}{2007}).

\bibitem[{\citenamefont{Adams et~al.}(2004)}]{STAR2}
\bibinfo{author}{\bibfnamefont{J.}~\bibnamefont{Adams}} \bibnamefont{et~al.}
  (\bibinfo{collaboration}{STAR Collaboration}), \bibinfo{journal}{Phys. Rev.
  Lett.} \textbf{\bibinfo{volume}{92}}, \bibinfo{pages}{171801}
  (\bibinfo{year}{2004}).

\bibitem[{\citenamefont{Arsene et~al.}(2007)}]{BRAHMS}
\bibinfo{author}{\bibfnamefont{I.}~\bibnamefont{Arsene}} \bibnamefont{et~al.}
  (\bibinfo{collaboration}{BRAHMS Collaboration}),
  \bibinfo{journal}{Phys.Rev.Lett.} \textbf{\bibinfo{volume}{98}},
  \bibinfo{pages}{252001} (\bibinfo{year}{2007}).

\bibitem[{\citenamefont{de~Florian
  et~al.}(2007{\natexlab{a}})\citenamefont{de~Florian, Vogelsang, and
  Wagner}}]{Resum}
\bibinfo{author}{\bibfnamefont{D.}~\bibnamefont{de~Florian}},
  \bibinfo{author}{\bibfnamefont{W.}~\bibnamefont{Vogelsang}},
  \bibnamefont{and} \bibinfo{author}{\bibfnamefont{F.}~\bibnamefont{Wagner}},
  \bibinfo{journal}{Phys. Rev. D.} \textbf{\bibinfo{volume}{76}},
  \bibinfo{pages}{094021} (\bibinfo{year}{2007}{\natexlab{a}}).

\bibitem[{\citenamefont{Almeida et~al.}(2009)\citenamefont{Almeida, Sterman,
  and Vogelsang}}]{Resum2}
\bibinfo{author}{\bibfnamefont{L.~G.} \bibnamefont{Almeida}},
  \bibinfo{author}{\bibfnamefont{G.~F.} \bibnamefont{Sterman}},
  \bibnamefont{and}
  \bibinfo{author}{\bibfnamefont{W.}~\bibnamefont{Vogelsang}},
  \bibinfo{journal}{Phys. Rev. D} \textbf{\bibinfo{volume}{80}},
  \bibinfo{pages}{074016} (\bibinfo{year}{2009}).

\bibitem[{\citenamefont{Adare et~al.}(2011{\natexlab{b}})}]{PH62idch}
\bibinfo{author}{\bibfnamefont{A.}~\bibnamefont{Adare}} \bibnamefont{et~al.}
  (\bibinfo{collaboration}{PHENIX Collaboration}), \bibinfo{journal}{Phys. Rev.
  C} \textbf{\bibinfo{volume}{83}}, \bibinfo{pages}{064903}
  (\bibinfo{year}{2011}{\natexlab{b}}).

\bibitem[{\citenamefont{Adler et~al.}(2008)}]{PHENIX_heavydAu}
\bibinfo{author}{\bibfnamefont{S.~S.} \bibnamefont{Adler}} \bibnamefont{et~al.}
  (\bibinfo{collaboration}{PHENIX Collaboration}), \bibinfo{journal}{Phys. Rev.
  C} \textbf{\bibinfo{volume}{77}}, \bibinfo{pages}{014905}
  (\bibinfo{year}{2008}).

\bibitem[{\citenamefont{Adler et~al.}(2004)}]{PHENIX_heavyAuAu}
\bibinfo{author}{\bibfnamefont{S.~S.} \bibnamefont{Adler}} \bibnamefont{et~al.}
  (\bibinfo{collaboration}{PHENIX Collaboration}), \bibinfo{journal}{Phys. Rev.
  C} \textbf{\bibinfo{volume}{69}}, \bibinfo{pages}{034910}
  (\bibinfo{year}{2004}).

\bibitem[{\citenamefont{Alekseev et~al.}(2003)}]{RHIC_polpp}
\bibinfo{author}{\bibfnamefont{I.}~\bibnamefont{Alekseev}}
  \bibnamefont{et~al.}, \bibinfo{journal}{Nucl. Instrum. Methods A}
  \textbf{\bibinfo{volume}{499}}, \bibinfo{pages}{392} (\bibinfo{year}{2003}).

\bibitem[{\citenamefont{Bunce et~al.}(2000)\citenamefont{Bunce, Saito, Soffer,
  and Vogelsang}}]{BunceAnnRev}
\bibinfo{author}{\bibfnamefont{G.}~\bibnamefont{Bunce}},
  \bibinfo{author}{\bibfnamefont{N.}~\bibnamefont{Saito}},
  \bibinfo{author}{\bibfnamefont{J.}~\bibnamefont{Soffer}}, \bibnamefont{and}
  \bibinfo{author}{\bibfnamefont{W.}~\bibnamefont{Vogelsang}},
  \bibinfo{journal}{Ann. Rev. Nucl. Part. Sci.} \textbf{\bibinfo{volume}{50}},
  \bibinfo{pages}{525} (\bibinfo{year}{2000}).

\bibitem[{\citenamefont{Bass}(2005)}]{SPINRev}
\bibinfo{author}{\bibfnamefont{S.~D.} \bibnamefont{Bass}},
  \bibinfo{journal}{Rev. Mod. Phys.} \textbf{\bibinfo{volume}{77}},
  \bibinfo{pages}{1257} (\bibinfo{year}{2005}).

\bibitem[{\citenamefont{Ashman et~al.}(1988)}]{EMC}
\bibinfo{author}{\bibfnamefont{J.}~\bibnamefont{Ashman}} \bibnamefont{et~al.}
  (\bibinfo{collaboration}{European Muon Collaboration}),
  \bibinfo{journal}{Phys. Lett. B} \textbf{\bibinfo{volume}{206}},
  \bibinfo{pages}{364} (\bibinfo{year}{1988}).

\bibitem[{\citenamefont{Anthony et~al.}(1996)}]{E142n}
\bibinfo{author}{\bibfnamefont{P.}~\bibnamefont{Anthony}} \bibnamefont{et~al.}
  (\bibinfo{collaboration}{E142 Collaboration}), \bibinfo{journal}{Phys. Rev.
  D} \textbf{\bibinfo{volume}{54}}, \bibinfo{pages}{6620}
  (\bibinfo{year}{1996}).

\bibitem[{\citenamefont{Abe et~al.}(1997)}]{E154n}
\bibinfo{author}{\bibfnamefont{K.}~\bibnamefont{Abe}} \bibnamefont{et~al.}
  (\bibinfo{collaboration}{E154 Collaboration}), \bibinfo{journal}{Phys. Rev.
  Lett.} \textbf{\bibinfo{volume}{79}}, \bibinfo{pages}{26}
  (\bibinfo{year}{1997}).

\bibitem[{\citenamefont{Ackerstaff et~al.}(1997)}]{HERMESn}
\bibinfo{author}{\bibfnamefont{K.}~\bibnamefont{Ackerstaff}}
  \bibnamefont{et~al.} (\bibinfo{collaboration}{HERMES Collaboration}),
  \bibinfo{journal}{Phys. Lett. B} \textbf{\bibinfo{volume}{404}},
  \bibinfo{pages}{383} (\bibinfo{year}{1997}).

\bibitem[{\citenamefont{Abe et~al.}(1998)}]{E143pd}
\bibinfo{author}{\bibfnamefont{K.}~\bibnamefont{Abe}} \bibnamefont{et~al.}
  (\bibinfo{collaboration}{E143 Collaboration}), \bibinfo{journal}{Phys. Rev.
  D} \textbf{\bibinfo{volume}{58}}, \bibinfo{pages}{112003}
  (\bibinfo{year}{1998}).

\bibitem[{\citenamefont{Adeva et~al.}(1998)}]{SMCpd}
\bibinfo{author}{\bibfnamefont{B.}~\bibnamefont{Adeva}} \bibnamefont{et~al.}
  (\bibinfo{collaboration}{Spin Muon Collaboration}), \bibinfo{journal}{Phys.
  Rev. D} \textbf{\bibinfo{volume}{58}}, \bibinfo{pages}{112001}
  (\bibinfo{year}{1998}).

\bibitem[{\citenamefont{Anthony et~al.}(2000)}]{E155p}
\bibinfo{author}{\bibfnamefont{P.}~\bibnamefont{Anthony}} \bibnamefont{et~al.}
  (\bibinfo{collaboration}{E155 Collaboration}), \bibinfo{journal}{Phys. Lett.
  B} \textbf{\bibinfo{volume}{493}}, \bibinfo{pages}{19}
  (\bibinfo{year}{2000}).

\bibitem[{\citenamefont{Zheng et~al.}(2004)}]{JLABn}
\bibinfo{author}{\bibfnamefont{X.}~\bibnamefont{Zheng}} \bibnamefont{et~al.}
  (\bibinfo{collaboration}{JLab Hall A Collaboration}), \bibinfo{journal}{Phys.
  Rev. C} \textbf{\bibinfo{volume}{70}}, \bibinfo{pages}{065207}
  (\bibinfo{year}{2004}).

\bibitem[{\citenamefont{Dharmawardane et~al.}(2006)}]{JLABp}
\bibinfo{author}{\bibfnamefont{K.}~\bibnamefont{Dharmawardane}}
  \bibnamefont{et~al.} (\bibinfo{collaboration}{CLAS Collaboration}),
  \bibinfo{journal}{Phys. Lett. B} \textbf{\bibinfo{volume}{641}},
  \bibinfo{pages}{11} (\bibinfo{year}{2006}).

\bibitem[{\citenamefont{Alexakhin et~al.}(2007)}]{COMPASSd}
\bibinfo{author}{\bibfnamefont{V.}~\bibnamefont{Alexakhin}}
  \bibnamefont{et~al.} (\bibinfo{collaboration}{COMPASS Collaboration}),
  \bibinfo{journal}{Phys. Lett. B} \textbf{\bibinfo{volume}{647}},
  \bibinfo{pages}{8} (\bibinfo{year}{2007}).

\bibitem[{\citenamefont{Airapetian et~al.}(2007)}]{HERMESpd}
\bibinfo{author}{\bibfnamefont{A.}~\bibnamefont{Airapetian}}
  \bibnamefont{et~al.} (\bibinfo{collaboration}{HERMES Collaboration}),
  \bibinfo{journal}{Phys. Rev. D} \textbf{\bibinfo{volume}{75}},
  \bibinfo{pages}{012007} (\bibinfo{year}{2007}).

\bibitem[{\citenamefont{Gl{\"{u}}ck et~al.}(2001)\citenamefont{Gl{\"{u}}ck,
  Reya, Stratmann, and Vogelsang}}]{GRSV}
\bibinfo{author}{\bibfnamefont{M.}~\bibnamefont{Gl{\"{u}}ck}},
  \bibinfo{author}{\bibfnamefont{E.}~\bibnamefont{Reya}},
  \bibinfo{author}{\bibfnamefont{M.}~\bibnamefont{Stratmann}},
  \bibnamefont{and}
  \bibinfo{author}{\bibfnamefont{W.}~\bibnamefont{Vogelsang}},
  \bibinfo{journal}{Phys. Rev. D.} \textbf{\bibinfo{volume}{63}},
  \bibinfo{pages}{094005} (\bibinfo{year}{2001}).

\bibitem[{\citenamefont{Bl{\"{u}}mlein and B{\"{o}}ttcher}(2010)}]{BB}
\bibinfo{author}{\bibfnamefont{J.}~\bibnamefont{Bl{\"{u}}mlein}}
  \bibnamefont{and}
  \bibinfo{author}{\bibfnamefont{H.}~\bibnamefont{B{\"{o}}ttcher}},
  \bibinfo{journal}{Nucl. Phys. B} \textbf{\bibinfo{volume}{841}},
  \bibinfo{pages}{205} (\bibinfo{year}{2010}).

\bibitem[{\citenamefont{Leader et~al.}(2010)\citenamefont{Leader, Sidorov, and
  Stamenov}}]{LSS2010}
\bibinfo{author}{\bibfnamefont{E.}~\bibnamefont{Leader}},
  \bibinfo{author}{\bibfnamefont{A.}~\bibnamefont{Sidorov}}, \bibnamefont{and}
  \bibinfo{author}{\bibfnamefont{D.}~\bibnamefont{Stamenov}},
  \bibinfo{journal}{Phys. Rev. D} \textbf{\bibinfo{volume}{82}},
  \bibinfo{pages}{114018} (\bibinfo{year}{2010}).

\bibitem[{\citenamefont{Vogelsang}()}]{subprocess}
\bibinfo{author}{\bibfnamefont{W.}~\bibnamefont{Vogelsang}},
  \bibinfo{note}{private communication (2008)}.

\bibitem[{\citenamefont{de~Florian et~al.}(2008)\citenamefont{de~Florian,
  Sassot, Stratmann, and Vogelsang}}]{DSSV}
\bibinfo{author}{\bibfnamefont{D.}~\bibnamefont{de~Florian}},
  \bibinfo{author}{\bibfnamefont{R.}~\bibnamefont{Sassot}},
  \bibinfo{author}{\bibfnamefont{M.}~\bibnamefont{Stratmann}},
  \bibnamefont{and}
  \bibinfo{author}{\bibfnamefont{W.}~\bibnamefont{Vogelsang}},
  \bibinfo{journal}{Phys. Rev. Lett.} \textbf{\bibinfo{volume}{101}},
  \bibinfo{pages}{072001} (\bibinfo{year}{2008}).

\bibitem[{\citenamefont{de~Florian
  et~al.}(2009{\natexlab{a}})\citenamefont{de~Florian, Sassot, Stratmann, and
  Vogelsang}}]{DSSV_prd}
\bibinfo{author}{\bibfnamefont{D.}~\bibnamefont{de~Florian}},
  \bibinfo{author}{\bibfnamefont{R.}~\bibnamefont{Sassot}},
  \bibinfo{author}{\bibfnamefont{M.}~\bibnamefont{Stratmann}},
  \bibnamefont{and}
  \bibinfo{author}{\bibfnamefont{W.}~\bibnamefont{Vogelsang}},
  \bibinfo{journal}{Phys. Rev. D} \textbf{\bibinfo{volume}{80}},
  \bibinfo{pages}{034030} (\bibinfo{year}{2009}{\natexlab{a}}).

\bibitem[{\citenamefont{Adcox et~al.}(2003{\natexlab{a}})}]{NIMPH}
\bibinfo{author}{\bibfnamefont{K.}~\bibnamefont{Adcox}} \bibnamefont{et~al.}
  (\bibinfo{collaboration}{PHENIX Collaboration}), \bibinfo{journal}{Nucl.
  Instr. and Meth. A} \textbf{\bibinfo{volume}{499}}, \bibinfo{pages}{469}
  (\bibinfo{year}{2003}{\natexlab{a}}).

\bibitem[{\citenamefont{Adcox et~al.}(2003{\natexlab{b}})}]{NIMPHCA}
\bibinfo{author}{\bibfnamefont{K.}~\bibnamefont{Adcox}} \bibnamefont{et~al.}
  (\bibinfo{collaboration}{PHENIX Collaboration}), \bibinfo{journal}{Nucl.
  Instr. and Meth. A} \textbf{\bibinfo{volume}{499}}, \bibinfo{pages}{489}
  (\bibinfo{year}{2003}{\natexlab{b}}).

\bibitem[{\citenamefont{Adare et~al.}(2006)}]{PHENIXmomresnew}
\bibinfo{author}{\bibfnamefont{A.}~\bibnamefont{Adare}} \bibnamefont{et~al.}
  (\bibinfo{collaboration}{PHENIX Collaboration}), \bibinfo{journal}{Phys. Rev.
  Lett.} \textbf{\bibinfo{volume}{97}}, \bibinfo{pages}{252002}
  (\bibinfo{year}{2006}).

\bibitem[{\citenamefont{Adare et~al.}(2009{\natexlab{b}})}]{PI062}
\bibinfo{author}{\bibfnamefont{A.}~\bibnamefont{Adare}} \bibnamefont{et~al.}
  (\bibinfo{collaboration}{PHENIX Collaboration}), \bibinfo{journal}{Phys. Rev.
  D} \textbf{\bibinfo{volume}{79}}, \bibinfo{pages}{012003}
  (\bibinfo{year}{2009}{\natexlab{b}}).

\bibitem[{\citenamefont{Drees and White}(2010)}]{vernier1}
\bibinfo{author}{\bibfnamefont{A.}~\bibnamefont{Drees}} \bibnamefont{and}
  \bibinfo{author}{\bibfnamefont{S.}~\bibnamefont{White}},
  \bibinfo{journal}{Conf.Proc.} \textbf{\bibinfo{volume}{C100523}},
  \bibinfo{pages}{MOPEC013} (\bibinfo{year}{2010}).

\bibitem[{\citenamefont{Drees et~al.}(2003)\citenamefont{Drees, Fox, Xu, and
  Huang}}]{vernier2}
\bibinfo{author}{\bibfnamefont{A.}~\bibnamefont{Drees}},
  \bibinfo{author}{\bibfnamefont{B.}~\bibnamefont{Fox}},
  \bibinfo{author}{\bibfnamefont{Z.}~\bibnamefont{Xu}}, \bibnamefont{and}
  \bibinfo{author}{\bibfnamefont{H.}~\bibnamefont{Huang}},
  \bibinfo{journal}{Conf.Proc.} \textbf{\bibinfo{volume}{C030512}},
  \bibinfo{pages}{1688} (\bibinfo{year}{2003}).

\bibitem[{\citenamefont{Nakagawa et~al.}(2008)}]{RHICpol}
\bibinfo{author}{\bibfnamefont{I.}~\bibnamefont{Nakagawa}}
  \bibnamefont{et~al.}, \bibinfo{journal}{AIP Conf.Proc.}
  \textbf{\bibinfo{volume}{980}}, \bibinfo{pages}{380} (\bibinfo{year}{2008}).

\bibitem[{\citenamefont{Okada et~al.}(2006)}]{HJetpol}
\bibinfo{author}{\bibfnamefont{H.}~\bibnamefont{Okada}} \bibnamefont{et~al.},
  \bibinfo{journal}{Phys. Lett. B} \textbf{\bibinfo{volume}{638}},
  \bibinfo{pages}{450} (\bibinfo{year}{2006}).

\bibitem[{\citenamefont{Adler et~al.}(2005)}]{PHRun2ch}
\bibinfo{author}{\bibfnamefont{S.~S.} \bibnamefont{Adler}} \bibnamefont{et~al.}
  (\bibinfo{collaboration}{PHENIX Collaboration}), \bibinfo{journal}{Phys. Rev.
  Lett.} \textbf{\bibinfo{volume}{95}}, \bibinfo{pages}{202001}
  (\bibinfo{year}{2005}).

\bibitem[{\citenamefont{Alper et~al.}(1975)}]{ISRBSC}
\bibinfo{author}{\bibfnamefont{B.}~\bibnamefont{Alper}} \bibnamefont{et~al.}
  (\bibinfo{collaboration}{ISR-BSC Collaboration}), \bibinfo{journal}{Nucl.
  Phys. B} \textbf{\bibinfo{volume}{87}}, \bibinfo{pages}{19}
  (\bibinfo{year}{1975}).

\bibitem[{\citenamefont{Drijard et~al.}(1982)}]{ISR_lambda}
\bibinfo{author}{\bibfnamefont{D.}~\bibnamefont{Drijard}} \bibnamefont{et~al.}
  (\bibinfo{collaboration}{CERN-Dortmund-Heidelberg-Warsaw Collaboration}),
  \bibinfo{journal}{Zeit. Phys.} \textbf{\bibinfo{volume}{C12}},
  \bibinfo{pages}{217} (\bibinfo{year}{1982}).

\bibitem[{GEA(1993)}]{GEANT3}
\emph{\bibinfo{title}{GEANT 3.2.1}}, \bibinfo{organization}{CERN Program
  Library} (\bibinfo{year}{1993}),
  \bibinfo{note}{\url{http://wwwasdoc.web.cern.ch/wwwasdoc/pdfdir/geant.pdf}}.

\bibitem[{\citenamefont{de~Florian and Wagner}()}]{pQCDXsec}
\bibinfo{author}{\bibfnamefont{D.}~\bibnamefont{de~Florian}} \bibnamefont{and}
  \bibinfo{author}{\bibfnamefont{F.}~\bibnamefont{Wagner}},
  \bibinfo{note}{private communication (2010)}.

\bibitem[{\citenamefont{Martin et~al.}(2003)\citenamefont{Martin, Roberts,
  Stirling, and Thorne}}]{MRST2002}
\bibinfo{author}{\bibfnamefont{A.}~\bibnamefont{Martin}},
  \bibinfo{author}{\bibfnamefont{R.}~\bibnamefont{Roberts}},
  \bibinfo{author}{\bibfnamefont{W.}~\bibnamefont{Stirling}}, \bibnamefont{and}
  \bibinfo{author}{\bibfnamefont{R.}~\bibnamefont{Thorne}},
  \bibinfo{journal}{Eur.Phys.J.} \textbf{\bibinfo{volume}{C28}},
  \bibinfo{pages}{455} (\bibinfo{year}{2003}).

\bibitem[{\citenamefont{de~Florian
  et~al.}(2007{\natexlab{b}})\citenamefont{de~Florian, Sassot, and
  Stratmann}}]{DSSfrag2}
\bibinfo{author}{\bibfnamefont{D.}~\bibnamefont{de~Florian}},
  \bibinfo{author}{\bibfnamefont{R.}~\bibnamefont{Sassot}}, \bibnamefont{and}
  \bibinfo{author}{\bibfnamefont{M.}~\bibnamefont{Stratmann}},
  \bibinfo{journal}{Phys. Rev. D} \textbf{\bibinfo{volume}{76}},
  \bibinfo{pages}{074033} (\bibinfo{year}{2007}{\natexlab{b}}).

\bibitem[{\citenamefont{Adler et~al.}(2003)}]{PHENIX1}
\bibinfo{author}{\bibfnamefont{S.~S.} \bibnamefont{Adler}} \bibnamefont{et~al.}
  (\bibinfo{collaboration}{PHENIX Collaboration}), \bibinfo{journal}{Phys. Rev.
  Lett.} \textbf{\bibinfo{volume}{91}}, \bibinfo{pages}{072303}
  (\bibinfo{year}{2003}).

\bibitem[{\citenamefont{Adare et~al.}(2007)}]{PHENIX2}
\bibinfo{author}{\bibfnamefont{A.}~\bibnamefont{Adare}} \bibnamefont{et~al.}
  (\bibinfo{collaboration}{PHENIX Collaboration}), \bibinfo{journal}{Phys. Rev.
  D.} \textbf{\bibinfo{volume}{76}}, \bibinfo{pages}{051106}
  (\bibinfo{year}{2007}).

\bibitem[{\citenamefont{de~Florian
  et~al.}(2009{\natexlab{b}})\citenamefont{de~Florian, Sassot, and
  Stratmann}}]{DSSfrag}
\bibinfo{author}{\bibfnamefont{D.}~\bibnamefont{de~Florian}},
  \bibinfo{author}{\bibfnamefont{R.}~\bibnamefont{Sassot}}, \bibnamefont{and}
  \bibinfo{author}{\bibfnamefont{M.}~\bibnamefont{Stratmann}},
  \bibinfo{journal}{Phys. Rev. D} \textbf{\bibinfo{volume}{75}},
  \bibinfo{pages}{114010} (\bibinfo{year}{2009}{\natexlab{b}}).

\bibitem[{\citenamefont{Aidala et~al.}(2011)\citenamefont{Aidala, Ellinghaus,
  Sassot, Seele, and Stratmann}}]{Etafrag}
\bibinfo{author}{\bibfnamefont{C.~A.} \bibnamefont{Aidala}},
  \bibinfo{author}{\bibfnamefont{F.}~\bibnamefont{Ellinghaus}},
  \bibinfo{author}{\bibfnamefont{R.}~\bibnamefont{Sassot}},
  \bibinfo{author}{\bibfnamefont{J.~P.} \bibnamefont{Seele}}, \bibnamefont{and}
  \bibinfo{author}{\bibfnamefont{M.}~\bibnamefont{Stratmann}},
  \bibinfo{journal}{Phys. Rev. D} \textbf{\bibinfo{volume}{83}},
  \bibinfo{pages}{034002} (\bibinfo{year}{2011}).

\bibitem[{\citenamefont{Taneja}()}]{NLOALL}
\bibinfo{author}{\bibfnamefont{S.}~\bibnamefont{Taneja}},
  \bibinfo{note}{private communication (2010)}.

\bibitem[{\citenamefont{Pumplin et~al.}(2002)\citenamefont{Pumplin, Stump,
  Huston, Lai, Nadolsky et~al.}}]{CTEQ6}
\bibinfo{author}{\bibfnamefont{J.}~\bibnamefont{Pumplin}},
  \bibinfo{author}{\bibfnamefont{D.}~\bibnamefont{Stump}},
  \bibinfo{author}{\bibfnamefont{J.}~\bibnamefont{Huston}},
  \bibinfo{author}{\bibfnamefont{H.}~\bibnamefont{Lai}},
  \bibinfo{author}{\bibfnamefont{P.~M.} \bibnamefont{Nadolsky}},
  \bibnamefont{et~al.}, \bibinfo{journal}{JHEP}
  \textbf{\bibinfo{volume}{0207}}, \bibinfo{pages}{012} (\bibinfo{year}{2002}).

\end{thebibliography}
%===========================

%=====================
\end{document}